\documentclass[11pt]{article}
\usepackage{amsmath,amsthm,amssymb}
\usepackage{graphicx}
\usepackage{mathptmx,times}

\makeatletter
\numberwithin{equation}{section}
\renewcommand\section{\@startsection {section}{1}{\z@}%
  {-4.6ex \@plus -0.4ex \@minus -0.2ex}{2.4ex \@plus0.2ex}%
  {\normalfont\bfseries}}
\renewcommand\subsection{\@startsection{subsection}{2}{\z@}%
  {-3.2ex\@plus -0.2ex \@minus -0.2ex}{1.8ex \@plus0.2ex}%
  {\normalsize\it}}
\renewcommand\paragraph{\@startsection{paragraph}{4}{\z@}%
  {2.0ex \@plus 0.2ex \@minus 0.2ex}{-1em}%
  {\normalfont\normalsize\bfseries}}

\def\@maketitle{\null\vskip2.4em%
  \begin{center} \let\footnote\thanks \vskip3.4em%
  {\Large\@title \par}\vskip2.4em%
  {\lineskip 0.5em\begin{tabular}[t]{c}\@author\end{tabular}\par}\vskip1em%
  {\@date}%
  \end{center}%
  \par\vskip3.4em}
\long\def\@makecaption#1#2{\vskip\abovecaptionskip
  \sbox\@tempboxa{\small #1: #2\par}%
  \ifdim \wd\@tempboxa >\hsize \small #1: #2\par
  \else \global \@minipagefalse \hb@xt@\hsize{\hfil\box\@tempboxa\hfil}%
  \fi
  \vskip\belowcaptionskip}
\advance\textwidth 4mm
\advance\textheight 24mm
\advance\voffset -18mm
\def\Real{\mathbb{R}}
\def\sech{\mathop{\rm sech}\nolimits}
\def\half{{\textstyle\frac12}}
\allowdisplaybreaks
\let\eref=\eqref
\let\eqref=\undefined

\advance\hoffset -10mm

\begin{document}
\title{\sf Soliton interactions of the Kadomtsev-Petviashvili equation and 
generation of large-amplitude water waves}
\author{\it By Gino Biondini$^1$, Ken-ichi Maruno$^2$, 
Masayuki Oikawa$^3$ and Hidekazu Tsuji$^3$}
\date{ }
\maketitle

\hrule
\vskip\bigskipamount
\begingroup\noindent
We study the maximum wave amplitude produced by 
line-soliton interactions of the Kadomtsev-Petviashvili 
II (KPII) equation, and we discuss a mechanism of 
generation of large amplitude shallow water waves 
by multi-soliton interactions of KPII.
We also describe a method to predict the possible 
maximum wave amplitude from asymptotic data.  
Finally, we report on numerical simulations of 
multi-soliton complexes of the KPII equation which 
verify the robustness of all types of soliton 
interactions and web-like structure. 
\endgroup
\vskip\bigskipamount
\hrule
\par\bigskip

\section{Introduction}

The existence of waves of large height on the sea surface 
is a dangerous phenomenon \cite{Hamer,Kharif,Li,pelinovsky,Soomere-ferry}. 
Extreme waves occur much more frequently than it 
might be expected from surface wave statistics \cite{Kharif}. 
These extreme waves, which are particularly steep and may 
arise both in deep water and in shallow water, 
have a significant impact on the safety of people and 
infrastructure, and are responsible for the erosion of 
coastlines and sea bottoms and changes to the biological 
environment. 
Thus, understanding the physics of these extreme waves is 
an important task which may even contribute to save lives. 
%
Although several physical mechanisms of generation of
extreme waves in deep water have been studied, 
less is known for the situation of shallow water 
\cite{pelinovsky}. 

Waves in shallow water have been studied since 
the nineteenth century, of course.
It is well-known that, in the case of weak nonlinearity, 
for weakly two-dimensional cases (that is, for the cases 
that the scale of variation in the direction normal to the propagation direction is much longer than that in 
the propagation direction) the fundamental equations 
for the dynamics in shallow water may be reduced to the Kadomtsev-Petviashvili II (KPII) equation 
\cite{ablowitz,kp,SegurFinkel}. 
Since the KP equation is integrable, the theory of 
integrable system can be used to analyze wave dynamics 
in detail. 
In particular, Someere \textit{et al.}\ recently 
studied the amplitude of ordinary 2-soliton solutions 
of the KPII equation\cite{Peterson,Soomere,Soomere2}. 
They pointed out that the interaction of two solitary 
waves may be one mechanism of generation of extreme waves.
In the case of ordinary 2-soliton solutions, the maximum 
interaction height can be four times that of the 
incoming solitary waves
in the limit of
the resonant 
Y-shaped solution originally found by Miles 
\cite{Miles1,Miles2,Oikawa}. 
The interaction pattern of the Miles solution and ordinary 2-soliton solutions,
however, is stationary.
Therefore, it cannot describe how a large amplitude wave can be generated.
%
Recently, we studied a class of exact line soliton solutions of the
KPII equation and found non-stationary interaction patterns and 
web-like structures \cite{jpa2003}. 
Such solutions describe resonant line-soliton interactions which are
generalization of the Y-shaped resonant soliton solution. 
More general solutions were later characterized in
\cite{prl2007,jmp2006,mcs2007,jpa2008,jpa2004}.

Our motivation in this paper is severalfold:
(i)~we determine the maximum amplitude in interactions of 
line solitons of the KPII equation, 
(ii)~we propose a mechanism to explain the generation of waves of large height,
(iii)~we describe an algorithm to determine the maximum
amplitude from experimental data,
(iv)~we perform numerical simulations of multi-soliton interactions 
of the KPII equation, whose results confirm the 
robustness and stability of soliton interactions and web-like structure.

\section{The KPII equation and its soliton solutions}
\label{s:kpbackground}

Here we briefly review some essential results the KPII equation
\begin{equation}
(-4u_t + 6uu_x + u_{xxx})_x + 3u_{yy} = 0
\label{e:KP}
\end{equation}
that will be referred to through the rest of this work.
Hereafter, $u(x,y,t)$ represents the dimensionless wave height 
to leading order, and
subscripts $x,y,t$ denote partial differentiation.
It is well known that solutions of \eref{e:KP}
can be expressed via a tau function $\tau(x,y,t)$ as
\begin{equation}
u(x,y,t) = 2 \frac{\partial^2}{\partial x^2}\log \tau (x,y,t), 
\label{e:utau}
\end{equation}
where $\tau(x,y,t)$ satisfies Hirota's bilinear equation 
\cite{Hirota,satsuma}.
Solutions of Hirota's equation can be written in terms of the Wronskian
determinant \cite{Freeman,FreemanNimmo} 
\begin{equation}
\tau(x,y,t)=  \mathop{\rm Wr}(f_1,\cdots , f_N)= 
  \det( f_n^{(n'-1)} )_{1\le n,n'\le N}\,,
\label{e:tauWronskian}
\end{equation}
with $f_n^{(j)}=\partial^j f_n/\partial x^j$, 
and where $f_1,\dots,f_N$ are 
a set of linearly independent solutions of the linear system
\begin{equation}
\frac{\partial f}{\partial y}=
\frac{\partial^2 f}{\partial x^2},\,\quad 
\frac{\partial f}{\partial t}=
\frac{\partial^3 f}{\partial x^3}\,. 
\label{e:linearPDEsyst} 
\end{equation}
For example, ordinary $N$-soliton solutions are obtained by taking
$f_n = e^{\theta_{2n-1}} + e^{\theta_{2n}}$ for $n = 1,\dots,N$,
where
\begin{equation}
\theta_m(x,y,t) =- k_mx + k_m^2y -k_m^3t + \theta_{m;0}\,
\label{e:thetadef}
\end{equation}
for $m=1,\dots,2N$,
where the $4N$ parameters
$k_1<\cdots<k_{2N}$ and $\theta_{1;0},\dots,\theta_{2N;0}$ are real constants. 
For $N=1$, one obtains the single-soliton solution of KPII:
\begin{equation}
u_{i,j}(x,y,t)=
  \half a_{i,j}^2\sech^2\big[\half(\theta_i-\theta_j)\big]\,, 
\label{e:KPIIsoliton}
\end{equation}
where $i=1$ and $j=2$.
Equation~\eref{e:KPIIsoliton} is a traveling-wave solution, 
[with wavenumber
$\mathbf{k}=(k_j-k_i,k_i^2-k_j^2)$ 
and frequency
$\omega=k_i^3-k_j^3$],
exponentially localized along the line $\theta_i=\theta_j$
of the $xy$-plane, and is therefore referred to as a \textit{line soliton}.
We refer to
\begin{equation}
a_{i,j}= k_j-k_i\,,\qquad d_{i,j}= k_i+k_j\,
\label{e:acdef}
\end{equation}
respectively as the soliton \textit{amplitude} and \textit{direction}.
Note that $d_{i,j}=\tan\alpha_{i,j}$, where 
$\alpha_{i,j}$ is the angle made by the soliton with the positive $y$-axis 
(counted clockwise), since
$\theta_i-\theta_j= (k_i-k_j)[-x + d_{i,j}y - (k_i^2+k_ik_j+k_j^2)t ]
 + \theta_{i;0} - \theta_{j;0}$.
The actual maximum of the solution, or wave height, is given by
$U_{i,j}= \frac12 a_{i,j}^2$.

\begin{figure}[t!]
\kern-1.5\bigskipamount
\begin{center}
\includegraphics[width=42.5mm,clip]{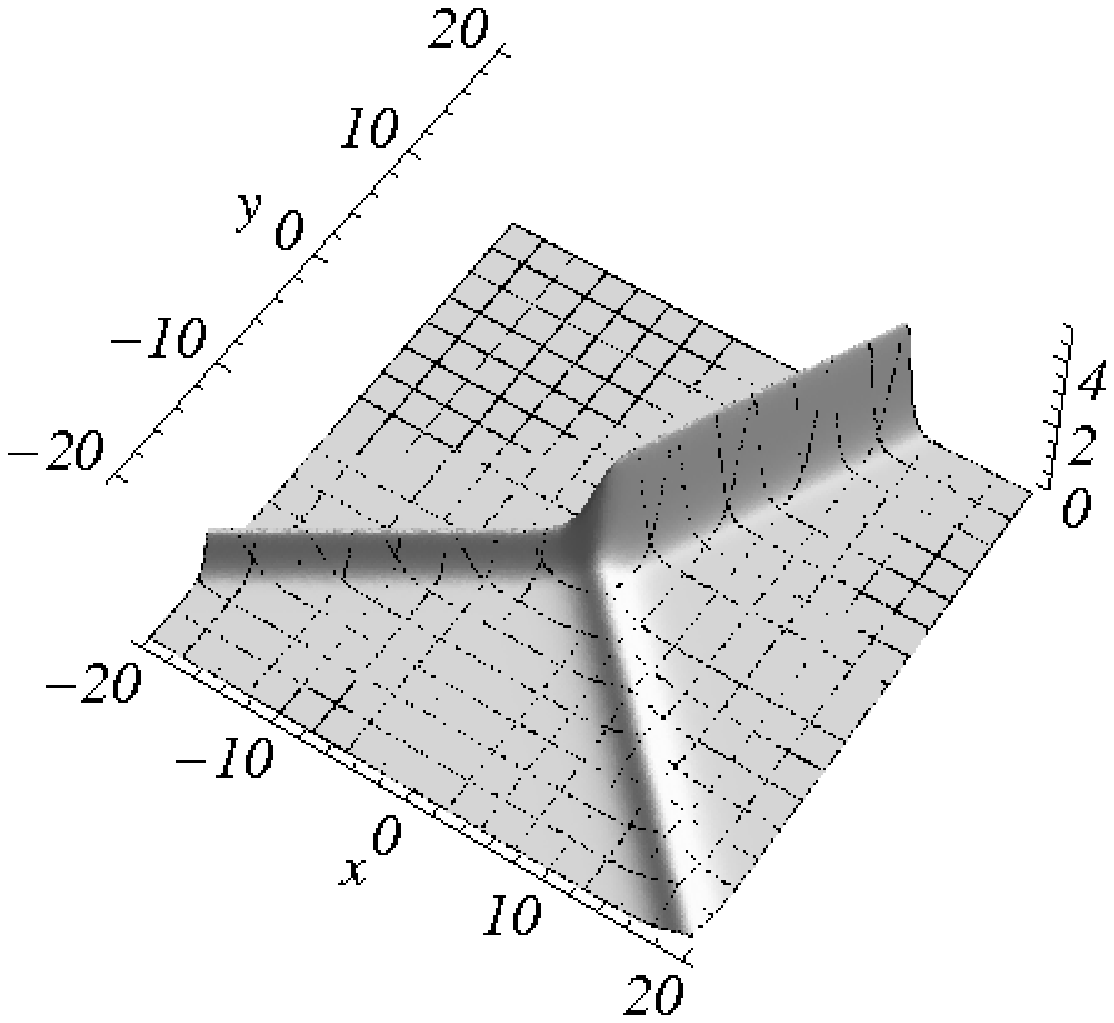}%
\includegraphics[width=42.5mm,clip]{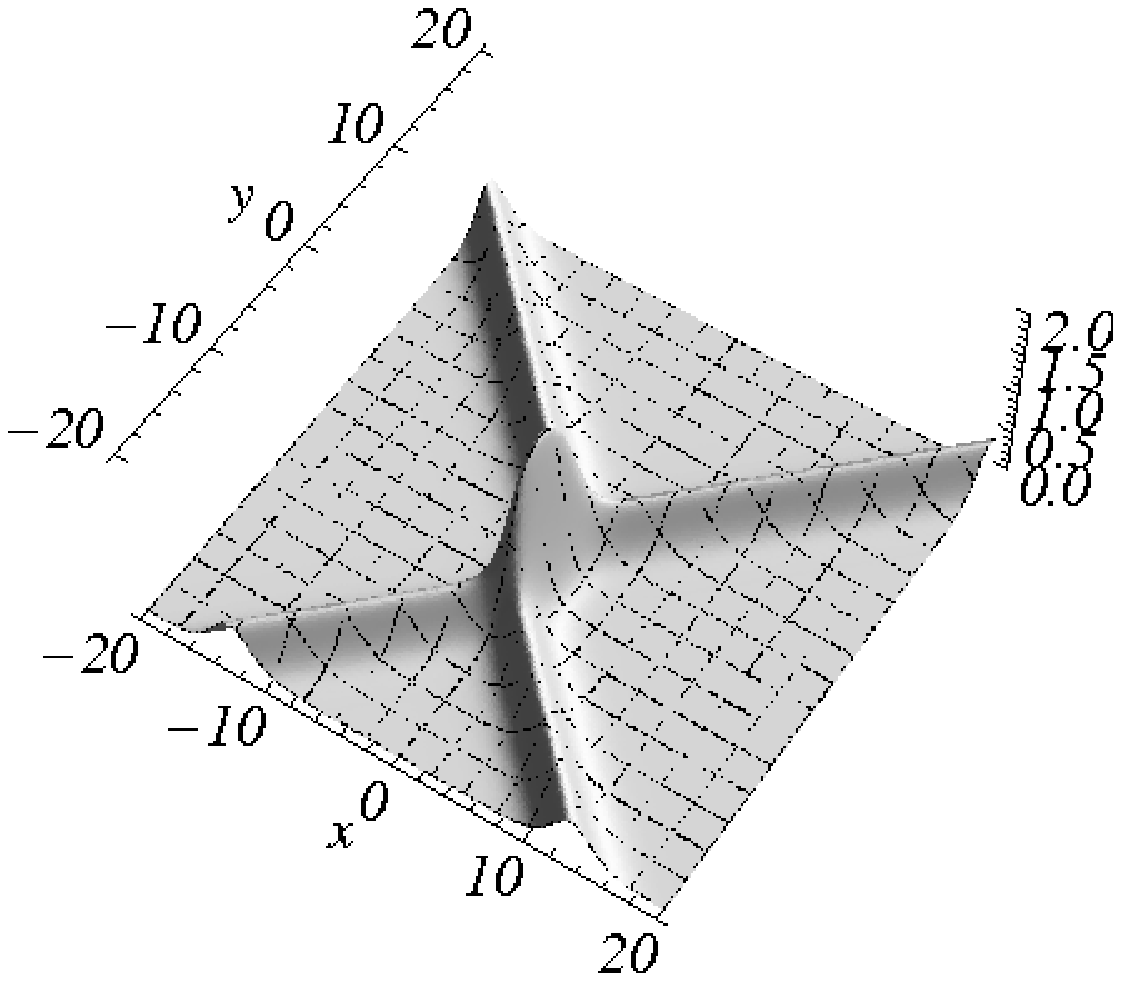}%
\includegraphics[width=42.5mm,clip]{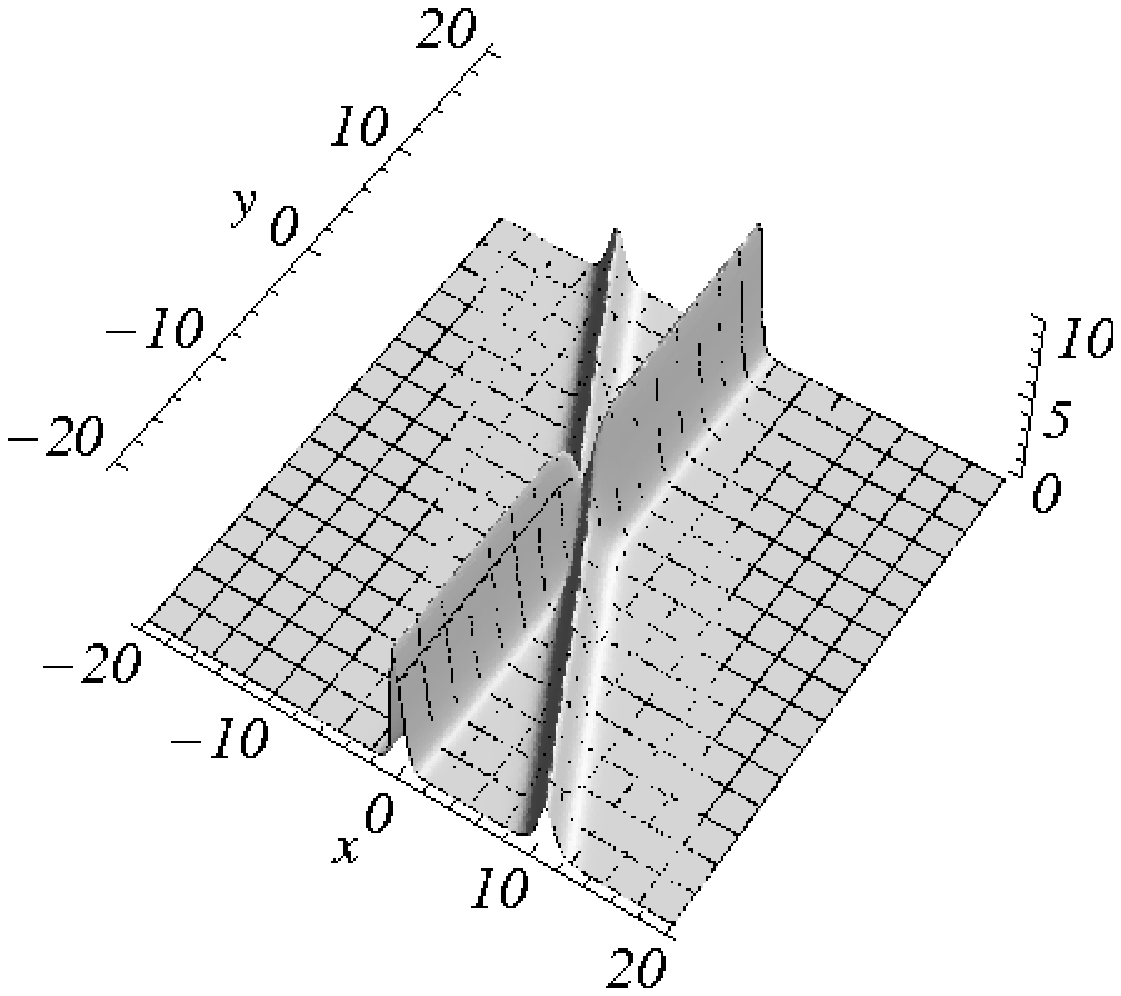}%
\end{center}
\caption{Left: Miles resonance, with 
$(k_1,k_2,k_3)= (-1,0,1.5)$
and $t=0$.
Center: ordinary 2-soliton solution, with 
$(k_1,\dots,k_4)=(-1,-0.001,0,1)$ 
and $t=0$.
Right: asymmetric 2-soliton solution, with
$(k_1,\dots,k_4)=(-2,-1.5,1,2)$ 
and $t=0$.
In all cases $\theta_{m;0}=0$ for $m=1,\dots,M$.}
\label{f:miles}
\kern-1.1\bigskipamount
\begin{center}
\includegraphics[width=42.5mm,clip]{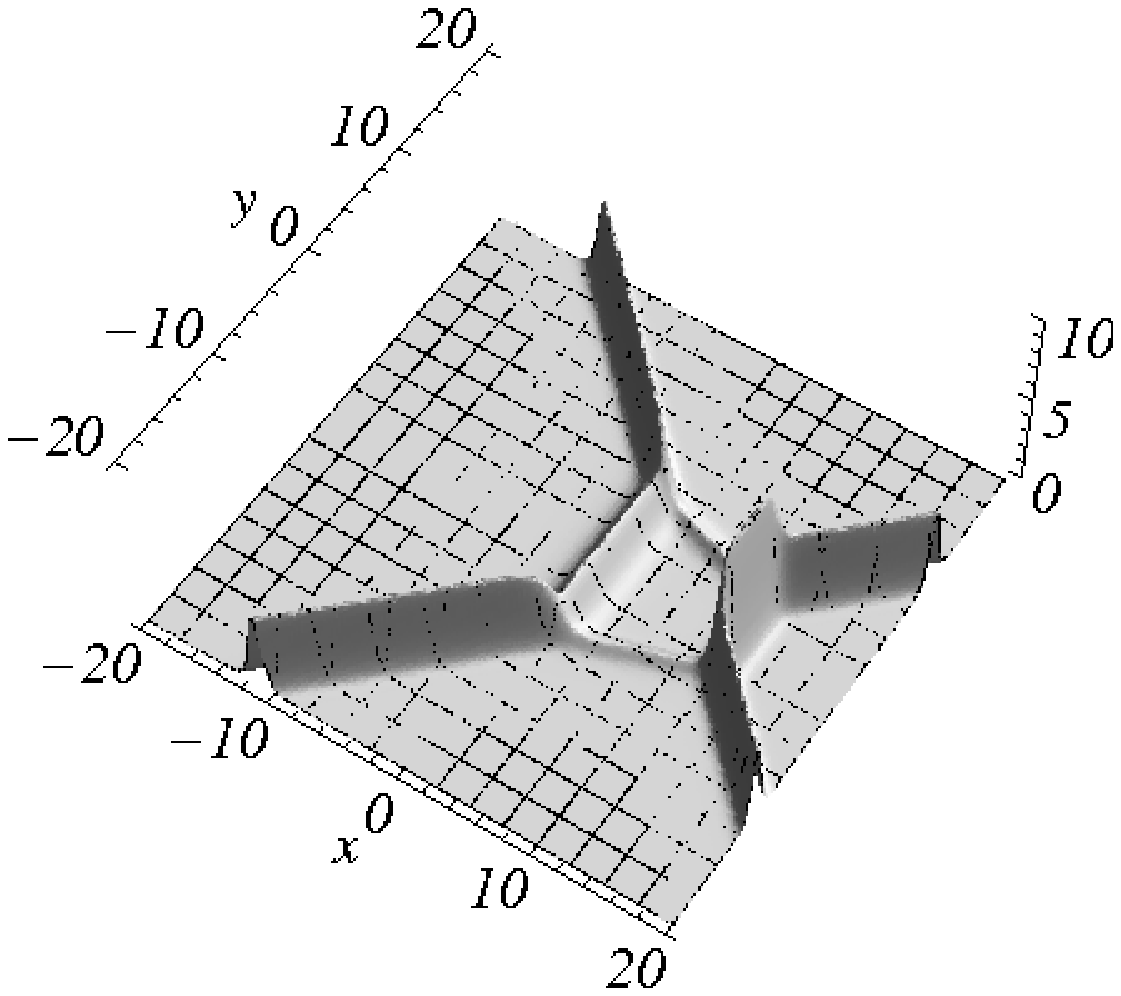}%
\includegraphics[width=42.5mm,clip]{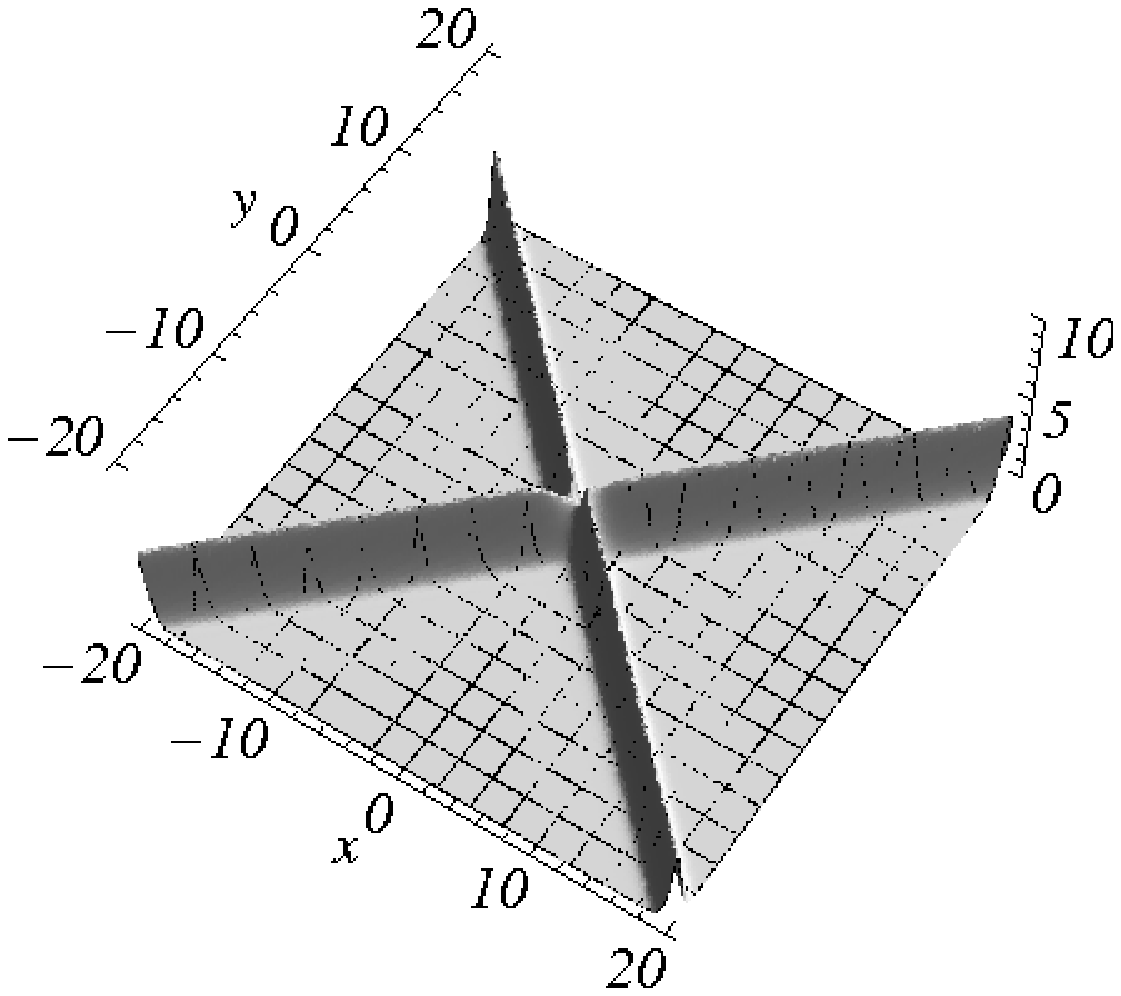}%
\includegraphics[width=42.5mm,clip]{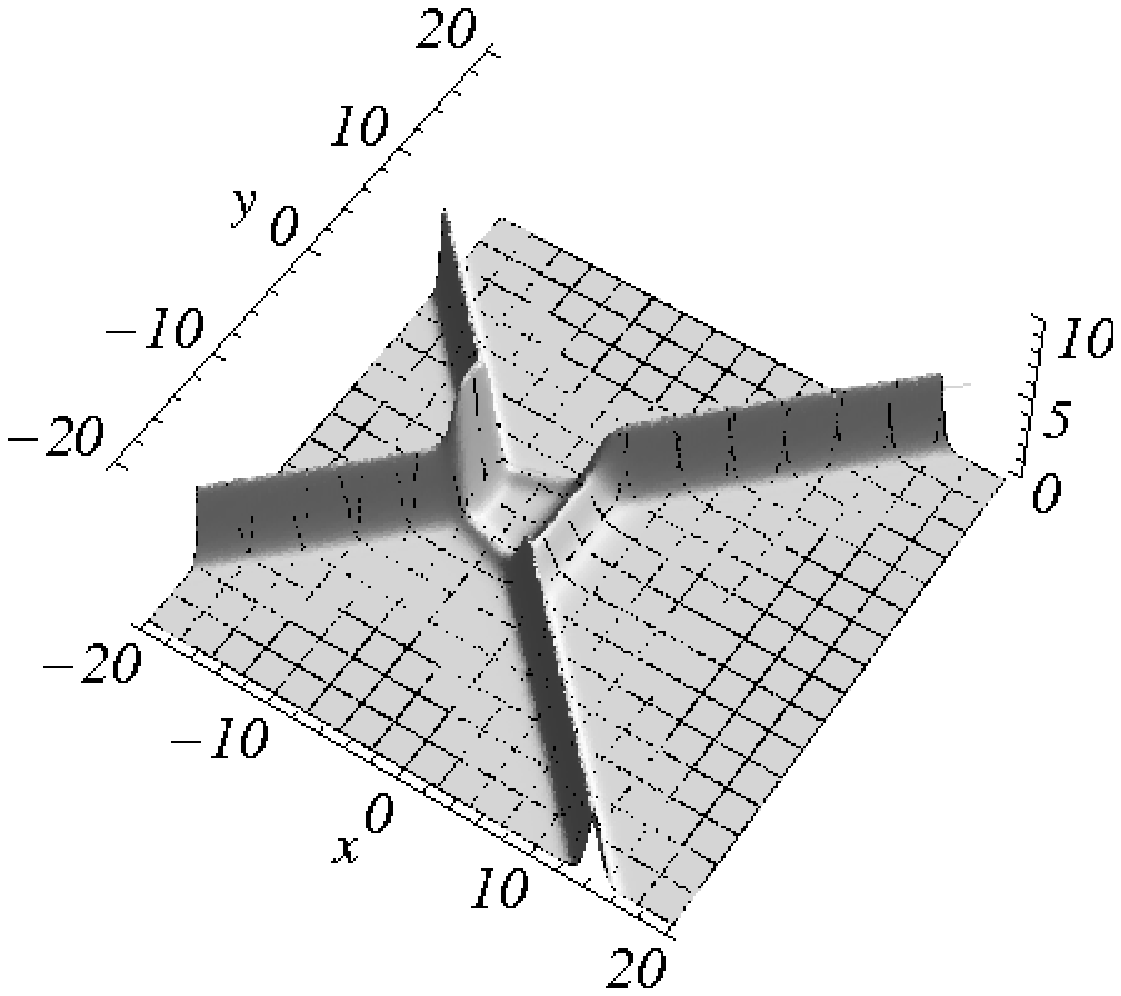}%
\end{center}
\caption{Time evolution of a resonant 2-soliton solution, with
$(k_1,\dots,k_4)=(-2,-1,1,2)$ and 
$\theta_{m;0}=0$ for $m=1,\dots,4$.
Left: $t=-3$; center: $t=0$; right: $t=2$.}
\label{f:resonant}
\end{figure}

The case with $N=1$ and $f= e^{\theta_1} + \cdots + e^{\theta_M}$,
with $\theta_1,\dots,\theta_M$ still given by~\eref{e:thetadef},
was studied in \cite{medina}.
In particular $M=3$ yields the Y-shaped solution often called 
\textit{Miles resonance},
in which three line solitons interact at a vertex, as shown in 
Fig.~\ref{f:miles},
and whose wavenumbers and phase parameters satisfy the resonance conditions
$\mathbf{k}_1+\mathbf{k}_2=\mathbf{k}_3$ and $\omega_1+\omega_2=\omega_3$.
(Note that, while the Miles resonance is also a traveling wave solutions, 
solutions with $M\ge4$ are not \cite{medina}.)
More in general, choosing $f_n= f^{(n-1)}$ for $n=1,\dots,N$ 
yields $\tau(x,y,t)$ in the form of a Hankel determinant.
In \cite{jpa2003} we studied such solutions with
$f= e^{\theta_1} + \cdots + e^{\theta_M}$, 
and we showed that they produce non-stationary and 
fully resonant $(N_-,N_+)$-soliton solutions of KPII,
that is, solutions with $N_-=M-N$ solitons asymptotically as $y \to -\infty$
and $N_+=N$ solitons asymptotically as $y \to \infty$,
and for intermediate values of $y$ these solitons interact resonantly,
i.e., via fundamental Miles resonances.

It should be clear that even more general solutions exist, however.
The most general linear combination of exponentials can be written as
\begin{equation}
f_n=\sum_{m=1}^M c_{n,m}\,e^{\theta_m}\,,
\label{e:fdef}
\end{equation}
with $\theta_m$ given by~\eref{e:thetadef} as before.
Then 
\eref{e:tauWronskian} 
is expanded as a sum of exponentials:
\begin{equation}
\tau(x,y,t)= \det(C\Theta K)=
  \sum_{1\le m_1<\cdots<m_N\le M}
    V_{m_1,\ldots,m_N}\,C_{m_1,\ldots,m_N}\,
  \exp\,\theta_{m_1,\cdots,m_N}\,,
\label{e:taugeneral}
\end{equation}
where 
$C= (c_{n,m})$ is the $N\times M$ coefficient matrix,
$\Theta={\rm diag}(e^{\theta_1},\cdots ,e^{\theta_M})$, 
and the $M\times N$ matrix $K$ is given by $K=(k_m^{n-1})$.
Hereafter, $\theta_{m_1,\cdots,m_N}$ denotes the phase combination
$\theta_{m_1,\cdots,m_N}(x,y,t)= 
\theta_{m_1} + \cdots + \theta_{m_N}\,$,
while $V_{m_1,\dots,m_N}$ is the Vandermonde determinant
$V_{m_1,\dots,m_N}= \prod_{1\le j<j'\le N}(k_{m_{j'}}-k_{m_j})\,$,
and $C_{m_1,\ldots,m_N}$ is the $N\times N$-minor whose $n$-th column 
is respectively given by the $m_n$-th column of the coefficient matrix
for $n = 1, \dots, N$.
The only time dependence in the tau function comes from the
exponential phases $\theta_{m_1,\dots,m_N}$.
Also, for all $G\in\mathrm{GL}(N,\Real)$,
the coefficient matrices $C$ and $C'= G\,C$
produce the same solution of KPII.
Thus without loss of generality one can consider $C$ to be in 
row-reduced echelon form (RREF),
which we will do throughout this work.
One can also multiply each column of $C$ by 
an arbitrary positive constant
which can be absorbed in the definition of $\theta_{1;0},\dots,\theta_{M;0}$.
\eject

Real nonsingular (positive) solutions of KPII are obtained if 
$k_1<\cdots<k_M$ and all minors of $C$ are nonnegative.
In \cite{jmp2006} we showed that,
under these assumptions and some fairly general irreducibility conditions 
on the coefficient matrix,
\eref{e:taugeneral} produces $(N_-,N_+)$-soliton solutions of KPII 
with $N_-=M-N$ and $N_+=N$, as in the simpler case of fully 
resonant solutions.
Asymptotic line solitons are given by~\eref{e:KPIIsoliton}
with the indices $i$ and $j$ labeling the phases $\theta_i$
and $\theta_j$ being swapped 
in the transition between two dominant phase combinations
along the line $\theta_i=\theta_j$.
Asymptotic solitons can thus be uniquely characterized by an index pair $[i,j]$
with $1\le i<j\le M$.
We call \textit{outgoing} line solitons those asymptotic as $y\to\infty$
and \textit{incoming} line solitons those asymptotic as $y\to-\infty$.
The index pairs are uniquely identified by appropriate rank conditions
on the minors of the coefficient matrix \cite{jmp2006}.
We also showed in \cite{jmp2006} that this decomposition is
time-independent, and that
the outgoing solitons are identified by pairs $[i_n^+,j_n^+]$,
$n=1,\dots,N$, where $i_1^+,\dots,i_N^+$ label the $N$ pivot columns of $C$;
similarly, the incoming solitons are identified by pairs $[i_n^-,j_n^-]$,
$n=1,\dots,M-N$, where $j_1^-,\dots,j_{M-N}^-$ label the $M-N$ non-pivots
columns of $C$.

In general, these solutions exhibit a mixture of resonant and 
non-resonant interaction patterns.
In the special case $M=2N$, leading to $N_-=N_+=N$,
we call \textit{elastic} $N$-soliton solutions those for which
the amplitudes and directions of the incoming solitons 
coincide with those of the outgoing solitons, while
all other $N$-soliton solutions are referred to as \textit{inelastic}.
Elastic $N$-soliton solutions were characterized in \cite{prl2007,jpa2004}.
The fully resonant line soliton solutions studied in \cite{jpa2003} 
are a special case of the tau function \eref{e:taugeneral}
in which all $N \times N$ minors of the coefficient matrix are nonzero.

\section{Interaction amplitudes of line-soliton solutions}
\label{s:interactions}

Here we first study the amplitude generated by 2-soliton 
interactions.           
These are obtained for $N=2$ and $M=4$, yielding the tau function
\begin{multline}
\tau(x,y,t)= 
  V_{1,2}C_{1,2}e^{\theta_{1,2}}
  + V_{1,3}C_{1,3}e^{\theta_{1,3}}
  + V_{1,4}C_{1,4}e^{\theta_{1,4}}
  + V_{2,3}C_{1,3}e^{\theta_{2,3}}
\\{ }
  + V_{2,4}C_{1,4}e^{\theta_{2,4}}
  + V_{3,4}C_{3,4}e^{\theta_{3,4}}\,,
\label{e:tau2soliton}
\end{multline}
where now $V_{i,j}= k_j-k_i$, and with $k_1<\cdots<k_4$ as before. 
The 2-soliton solutions of KPII were recently studied in 
several papers \cite{boiti,jpa2004,jmp2006,mcs2007,jpa2008,Chak-Kodama}.

\subsection{Elastic 2-soliton interactions}

It was shown in \cite{jpa2004} that elastic 2-soliton solutions 
are classified into three classes, 
shown in Figs.~\ref{f:miles} and~\ref{f:resonant}: 
ordinary (O-type), asymmetric (P-type) and resonant (T-type).
The coefficient matrices corresponding to these classes have 
the following RREFs:
\begin{gather}
C_{\rm O}=\left(\!\begin{array}{cccc}
1 & 1 & 0 & 0 \\
0 & 0 & 1 & 1 \\
\end{array}\!\right)\!,\!\quad 
C_{\rm P}=\left(\!\begin{array}{cccc}
1 & 0 & 0 & -1 \\
0 & 1 & 1 & 0 \\
\end{array}\!\right)\!,\!\quad 
C_{\rm T}=\left(\!\begin{array}{cccc}
1 & 0 & -1 & -1 \\
0 & 1 & c_{2,3} & c_{2,4} \\
\end{array}\!\right)\!,
\nonumber
\\
\label{e:Aelastic2soliton}
\end{gather}
with $c_{2,3}>c_{2,4}>0$.
These three types of solutions 
cover disjoint sectors of the 2-soliton parameter space of
amplitudes and directions
\cite{prl2007}.
Moreover, their interaction properties are also different.
This difference is obvious in the case of resonant solutions,
but also applies to ordinary and asymmetric solutions, since
asymmetric solutions only exist for unequal amplitude, and 
the interaction phase shift has the opposite sign for
ordinary and asymmetric solutions \cite{prl2007}.
We next show that these solutions also differ in terms of
the interaction amplitudes.

\paragraph{Ordinary 2-soliton interactions.}

Consider the coefficient matrix $C_{\rm O}$ in
\eref{e:Aelastic2soliton}.
It is $C_{1,2}=C_{3,4}=0$ in~\eref{e:tau2soliton}, 
while all other minors are unity.
In this case, the dominant phase combination as $x \to -\infty$ is $(2,4)$
[meaning that $e^{\theta_{2,4}}$ is the dominant exponential in~\eref{e:tau2soliton}], 
while that as $x \to \infty$ is $(1,3)$. 
The asymptotic solitons are [1,2] and [3,4]. 
The pattern of this solution is stationary. 
The maximum height was discussed in \cite
{Peterson,Soomere,Soomere2,Chak-Kodama}. 
It is expected that the value of $u(x,y,t)$ at the 
interaction center would give the global 
maximum. In fact, it was verified in \cite
{Soomere}. According to \cite{Chak-Kodama}, 
the value of $u(x,y,t)$ at the interaction center, $u_{\rm O}^{\rm (ic)}$,  
is given by 
\begin{equation}
u_{\rm O}^{\rm (ic)} = \frac{1}{2}(a_{1,2}^2 + a_{3,4}^2) 
+ a_{1,2}a_{3,4}\,\mbox{tanh}\left(\frac{\varDelta_{\rm O}}{4}\right),
\label{e:O-center-value}
\end{equation}
where
\begin{equation}
\varDelta_{\rm O} = \log\frac{(k_3-k_1)(k_4-k_2)}
{(k_3-k_2)(k_4-k_1)}=\log\frac{(
d_{3,4}
-
d_{1,2}
)^2-(a_{3,4}-a_{1,2})^2}
{(
d_{3,4}
-
d_{1,2}
)^2-(a_{3,4}+a_{1,2})^2}.
\label{e:O-Delta}
\end{equation}  
Since $(k_3-k_1)(k_4-k_2)-(k_3-k_2)(k_4-k_1)=
(k_2-k_1)(k_4-k_3) > 0$, $\varDelta_{\rm O} > 0$. So, 
\begin{equation}
\frac{1}{2}(a_{1,2}^2+a_{3,4}^2) < u_{\rm O}^{({\rm ic})} < 
\frac{1}{2}(a_{1,2}+a_{3,4})^2
\label{e:O-center-value-range}
\end{equation}
The left hand side is the infinitesimal limit 
$k_2-k_1 \to 0^+$ or $k_4-k_3 \to 0^+$ and is  
the sum of the heights of the asymptotic solitons 
$U_{1,2}$ and $U_{3,4}$. Accordingly, 
the maximum height $u_{\rm max}=u_{\rm O}^{({\rm ic})}$ 
is always greater than the sum of the heights of 
the asymptotic solitons.  
It is also lesser than $u_{\rm O}^{(\rm cr)}:=\frac12 (a_{1,2}+a_{3,4})^2$ as long as $k_1 < k_2 < k_3 < k_4$. 
$u_{\rm max}$ takes the critical value $u_{\rm O}^{(\rm cr)}$ 
in the limit $k_3-k_2 \to 0^+$.   
From $a_{1,2}a_{3,4} \le (a_{1,2}^2 + a_{3,4}^2)/2$, for $a_{1,2}=a_{3,4}=:a$, 
$u_{\rm O}^{({\rm cr})}$ takes its maximum $2a^2 = 4(a^2/2)$ 
which is four times the height of the incoming solitons. 
However, it should be noted that in the limit $k_3-k_2 \to 0^+$ 
the solution degenerates to a Y-shape resonant solution. 

It is also important to note that since $\frac12 (a_{1,2} + a_{3,4})^2  \le 4(\mbox{max}\{U_{1,2}, U_{3,4}\})$ as pointed out in \cite{Chak-Kodama}, the maximum height of $u(x,y,t)$ is less than four times the height of the taller(tallest) asymptotic soliton.  

The right hand side of~\eref{e:O-center-value-range} 
may be also obtained in the following way.
Near the critical angle ($k_2\approx k_3$),
the interaction arm appears in the intermediate region 
of 2 soliton interaction. The interaction arm corresponds 
to the double-phase transition $(1,3) \leftrightarrow 
(2,4)$ among the dominant phases in the $xy$-plane, 
which is located along the line $\theta_{1,3}=\theta_{2,4}$. 
This is due to the fact that the [1,2]-soliton in 
$y \to \infty$ is shifted to the right relative to that in 
$y \to -\infty$, while the [3,4]-soliton in $y \to \infty$ 
is shifted to the left relative to that in $y \to -\infty$. 
In a neighborhood of the line corresponding to 
the double-phase transition
$(i_1,i_2) \leftrightarrow (j_1,j_2)$, the tau function 
and the corresponding solution of KP are approximately, 
up to exponentially smaller terms,
\begin{gather}
\tau_{i_1i_2,j_1j_2}(x,y,t)= V_{i_1,i_2}e^{\theta_{i_1,i_2}}
  + 
V_{j_1,j_2}
 e^{\theta_{j_1,j_2}}\,,
\\
\noalign{\noindent and}
u_{i_1i_2,j_1j_2}(x,y,t)=
  \half a_{i_1i_2,j_1j_2}^2
    \sech^2\big[\half (\theta_{j_1,j_2}-\theta_{i_1,i_2}+
      \Delta_{i_1i_2,j_1j_2} )\big]\,, 
\label{e:doublesoliton}
\end{gather}
where $\Delta_{i_1i_2,j_1j_2}= \log(V_{j_1,j_2}/V_{i_1,i_2})$ 
and
\begin{equation}
a_{i_1i_2,j_1j_2}=k_{j_1}+k_{j_2}-(k_{i_1}+k_{i_2}).
\label{e:doublesolamplitude}
\end{equation}
The double-soliton~\eref{e:doublesoliton} 
is \textit{not}
in itself an exact solution of the KPII equation 
(and hence it is not a true line soliton),
because its wavevector and frequency do not satisfy the 
soliton dispersion relation~\cite{infeld}. 
This relation is satisfied in the limit $k_3-k_2\to0^+$. 
In this limit 
there are only three phases in the
tau function, and the solution degenerates to a Y-junction. 
In this limit, the interaction arm of 
ordinary 2-soliton solutions is given by
$u(x,y,t)= u_{13,24}(x,y,t), \ 
a_{13,24}=a_{1,2}+a_{3,4}$. 
This result agrees with the right hand side of
~\eref{e:O-center-value-range}. 
This method may be, therefore, effective for deriving  
the supremum for the height of the ordinary 2-soliton solution. 

\paragraph{Asymmetric 2-soliton interactions.}

Consider now $C_{\rm P}$ in \eref{e:Aelastic2soliton}.
It is $C_{1,4}=C_{2,3}=0$ while all other minors are unity.
The asymptotic solitons are [1,4] and [2,3], and
the dominant phase combinations as 
$x \to -\infty$ and $x \to \infty$ are $(3,4)$ and $(1,2)$, respectively. 
According to \cite{Chak-Kodama}, 
the value of $u(x,y,t)$ at the interaction center, $u_{\rm P}^{\rm (ic)}$,  
is given by 
\begin{equation}
u_{\rm P}^{\rm (ic)} = \frac{1}{2}(a_{1,4}^2 + a_{2,3}^2) 
- a_{1,4}a_{2,3}\,\mbox{tanh}\left(\frac{\varDelta_{\rm P}}{4}\right),
\label{e:P-center-value}
\end{equation}
where
\begin{equation}
\varDelta_{\rm P} = \log\frac{(k_3-k_1)(k_4-k_2)}
{(k_2-k_1)(k_4-k_3)}=\log\frac{(a_{1,4}+a_{2,3})^2-(
d_{1,4}
-
d_{2,3}
)^2}
{(a_{1,4}-a_{2,3})^2-(
d_{1,4}
-
d_{2,3}
)^2} > 0.
\label{e:P-Delta}
\end{equation} 
Accordingly, we have the estimate
\begin{equation}
\frac{1}{2}(a_{1,4}-a_{2,3})^2 < u_{\rm P}^{\rm (ic)} \le  
\frac{1}{2}a_{1,4}^2.
\label{e:P-center-value-range}
\end{equation}
The left hand side is the resonance limit $k_2-k_1 \to 0^+$ or 
$k_4 - k_3 \to 0^+$. The right hand side is the 
infinitesimal limit $k_3-k_2 \to 0^+$. Note that in this limit 
$a_{2,3} \to 0$. It is also noted that more accurate 
estimate for right hand side can be obtained (see \cite{Chak-Kodama}). 
Unlike the ordinary 2-soliton solution, $u_{\rm P}^{\rm (ic)}$ 
does not give the global maximum of $u(x,y,t)$. It is rather 
given by the height of the highest asymptotic soliton $U_{1,4}=
\frac12a_{1,4}^2$.  

The left hand side of \eref{e:P-center-value-range} can be also obtained in the same way as in the ordinary 2-soliton solution. 
For clarity, let us assume $k_1+k_4 > k_2+k_3$. In this case, 
[1,4]-soliton is located on the right side of [2,3]-soliton in 
$y \to \infty$ as in Fig.\ref{f:miles}. 
The [1,4]-soliton in $y \to \infty$ is shifted to the right 
relative to that in $y \to -\infty$, while the [2,3]-soliton 
in $y \to \infty$ is shifted to the left 
relative to that in $y \to -\infty$. Accordingly, the 
interaction arm appears as the boundary between the 
two dominant phase regions $(1,3)$ in $y \to -\infty$ and $(2,4)$ 
in $y \to \infty$.
Thus, the amplitude of the interaction
arm in the resonance limit $k_2-k_1 \to 0^+$ is given by 
$a_{13,24}$ as for ordinary solutions. 
Since $a_{13,24}=k_2+k_4-(k_1+k_3)=k_4-k_1-(k_3-k_2)=
a_{1,4}-a_{2,3}$, this result agrees with the left hand side 
of \eref{e:P-center-value-range}. For $k_1+k_4 < k_2+k_3$, 
we can also the same final result. 

For $k_1+k_4 = k_2+k_3$, [1,4] and [2,3] solitons are 
parallel and one-dimensional over-taking interaction 
takes place. Even for this case, \eref{e:P-center-value} 
is still valid if we take the interaction center 
as that in the $(x-(k_1+k_4)y, t)$-plane. According to 
\eref{e:P-center-value}, the height at the interaction 
center is $U_{1,4}-U_{2,3}$.


\paragraph{Resonant 2-soliton interactions.}

For the coefficient matrix $C_{\rm T}$ in~\eref{e:Aelastic2soliton}
all minors are nonzero.
The dominant phase combinations as $x\to -\infty$ and $x\to \infty$ 
are respectively $(3,4)$ and $(1,2)$, as with asymmetric solutions,
but the asymptotic solitons here are [1,3] and [2,4].
Here the interaction is mediated by four interaction segments 
and the solution is non-stationary.
(Its time evolution is shown in Fig.~\ref{f:resonant}.)
Hence, the situation appears at first to be more complicated than 
in the other two cases.
Nonetheless, the calculations are actually simpler,
because all the intermediate arms are true line solitons
of the KPII equation, 
and the resonance condition is satisfied at all vertices.
Moreover, the interaction pattern obeys the reflection symmetry
$(x,y,t)\mapsto(-x,-y,-t)$.
Indeed, it would be trivial (however, see the last 
part of this subsection) to see that, at all times, 
the tallest intermediate soliton is [1,4],  
which is obviously taller than either of the asymptotic solitons. 
Note, however, that 
$a_{1,3}+a_{2,4}=a_{1,4}+a_{2,3}>a_{1,4}$. 
Thus the maximum interaction amplitude is always less 
than the sum of the amplitudes of the asymptotic solitons.
In particular, in the case of equal-amplitude solitons, 
$a_{1,3}=a_{2,4}=:a$, we have $U_\mathrm{max}= \frac12a_{1,4}^2< 2a^2$. 
Thus the maximum height is less
than four times that of the asymptotic solitons.

Figure 2 shows that the interaction at $t=0$ neither generate 
the intermediate solitons nor large amplitude. At least, 
we can prove $u(0,0,0) < \frac12 a_{1,4}^2$ for u(x,y,t) 
with zero phase constants $\theta_{m;0}=0, m=1,\cdots,4$. 
Our method is as follows: We express $u(0,0,0)$,which is 
a function of $k_i$, in terms of $d_{i,j}$ and $a_{i,j}$. 
It turns out that the terms containing $d_{i,j}$ are 
in the form $(d_{i,j}-d_{m,n})^2 a_{i,j}a_{m,n}$. 
Accordingly, all the  
terms can be expressed in terms of $a_{i,j}$. 
If we make decomposition like $a_{1,4}=a_{1,2}+a_{2,3}+a_{3,4}$, 
we find that $\frac12 a_{1,4}^2-u(0,0,0)$ is a sum of 
positive terms.   
  
\subsection{Inelastic 2-soliton interactions}

Inelastic 2-soliton solutions fall into four categories \cite{jpa2008}, 
identified by the following coefficient matrices in RREF:
\begin{equation}
\begin{array}{cc}
\displaystyle 
C_{\rm I}= \left(\!\begin{array}{cccc}
  1 & 1 & 0 & -r \\
  0 & 0 & 1 & 1 \\
\end{array}\!\right)\,,\qquad
C_{\rm II}= \left(\!\begin{array}{cccc}
  1 & 0 & -r & -r \\
  0 & 1 & 1 & 1 \\
\end{array}\!\right)\,,
\\
\displaystyle
C_{\rm III}= \left(\!\begin{array}{cccc}
  1 & 0 & 0 & -r \\
  0 & 1 & 1 & 1 \\
\end{array}\!\right)\,,\qquad
C_{\rm IV}= \left(\!\begin{array}{cccc}
  1 & 0 & -r & -1 \\
  0 & 1 & 1 & 0 \\
\end{array}\!\right)\,,
\end{array}
\label{e:Ainelastic2s}
\end{equation}
with $r>0$.
We consider each of these in turn. 
For simplicity in what follows we set $r=1$. 
Note that in all of these cases exactly one of 
the minors of $C$ is zero
while all other minors are unity,
which produces a tau function 
with five phase combinations.

\begin{figure}[t!]
\kern-1.5\bigskipamount
\begin{center}
\includegraphics[width=42.5mm,clip]{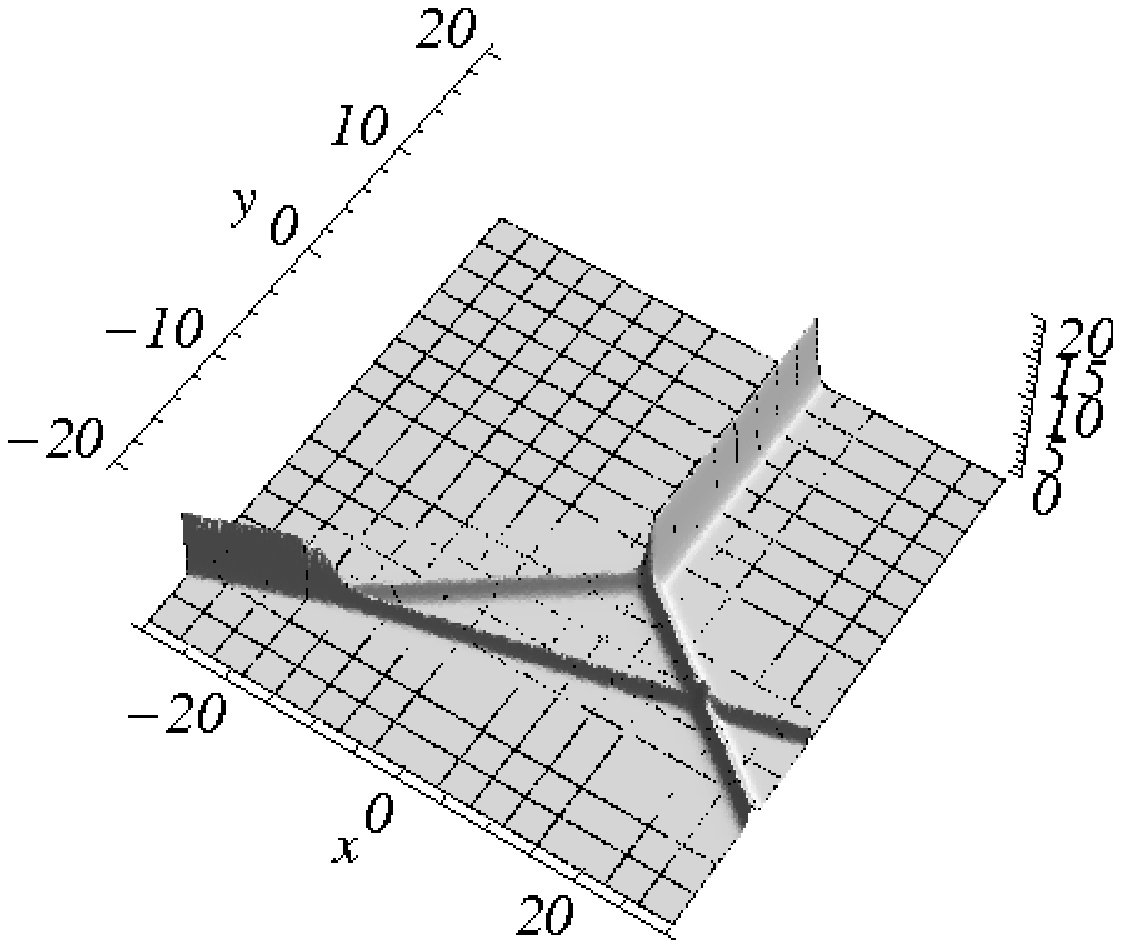}%
\includegraphics[width=42.5mm,clip]{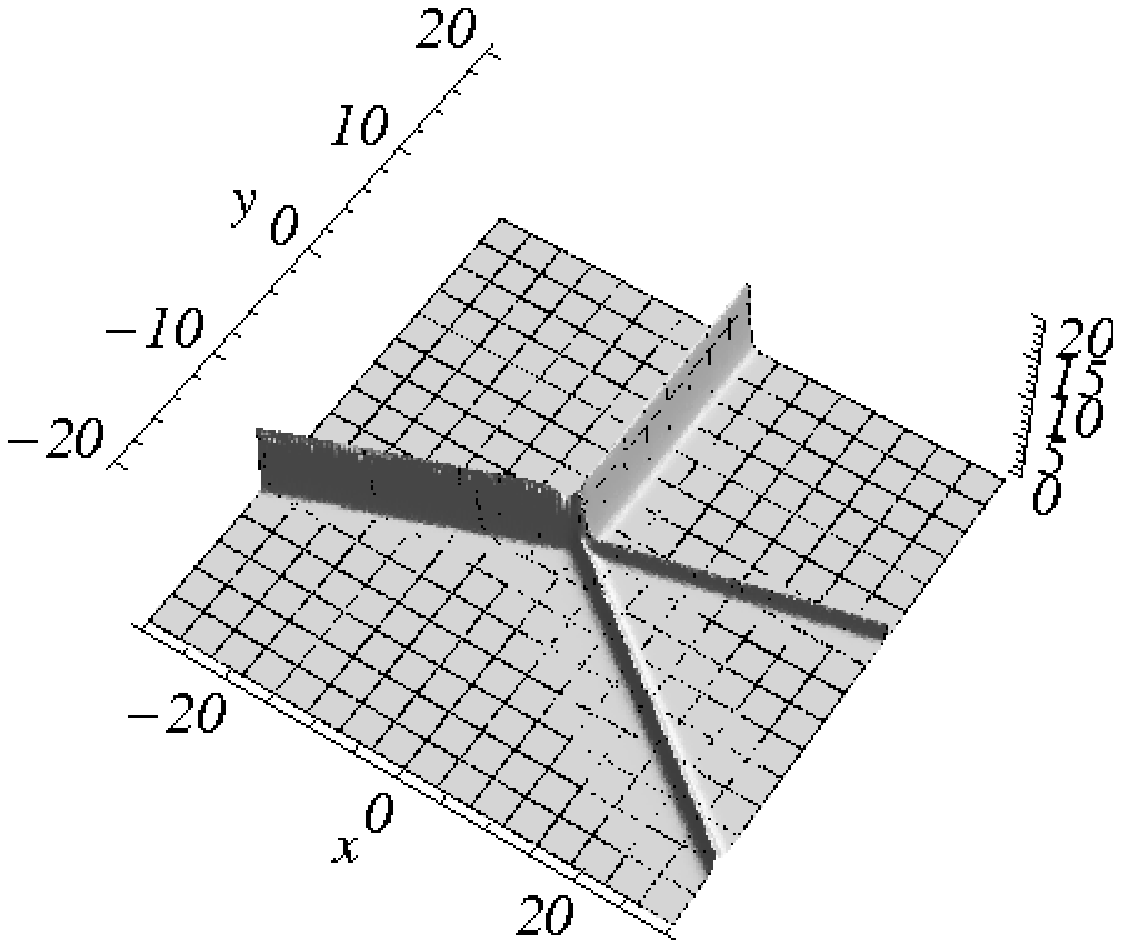}%
\includegraphics[width=42.5mm,clip]{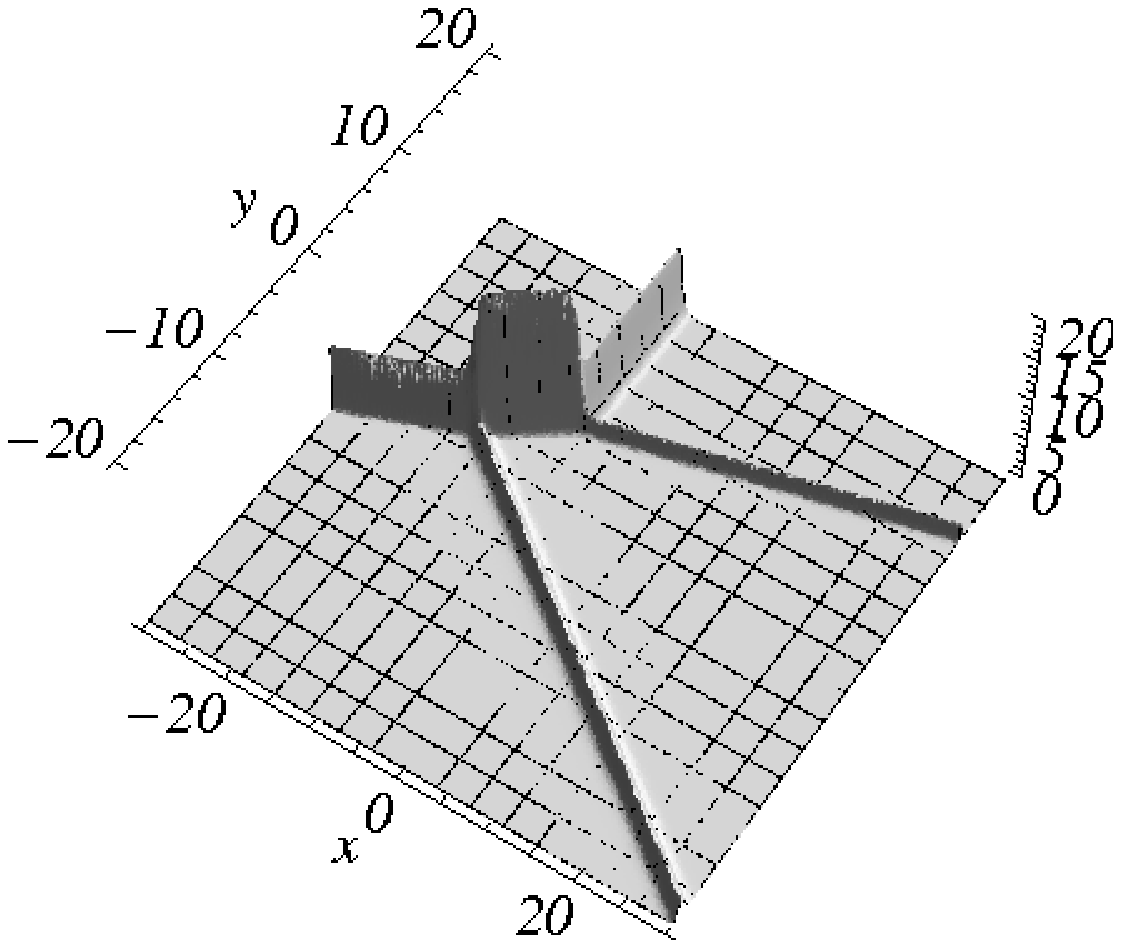}%
\end{center}
\smallskip
\caption{Time evolution of the inelastic 2-soliton solution obtained from
the coefficient matrix $C_\mathrm{I}$ in \eref{e:Ainelastic2s} with 
$(k_1,\dots,k_4)=(-2,0,2,4)$.
Left: $t=-2$; center: $t=0$; right: $t=2$.}
\label{f:inelasticIs}
\medskip
\begin{center}
\includegraphics[width=40mm,clip]{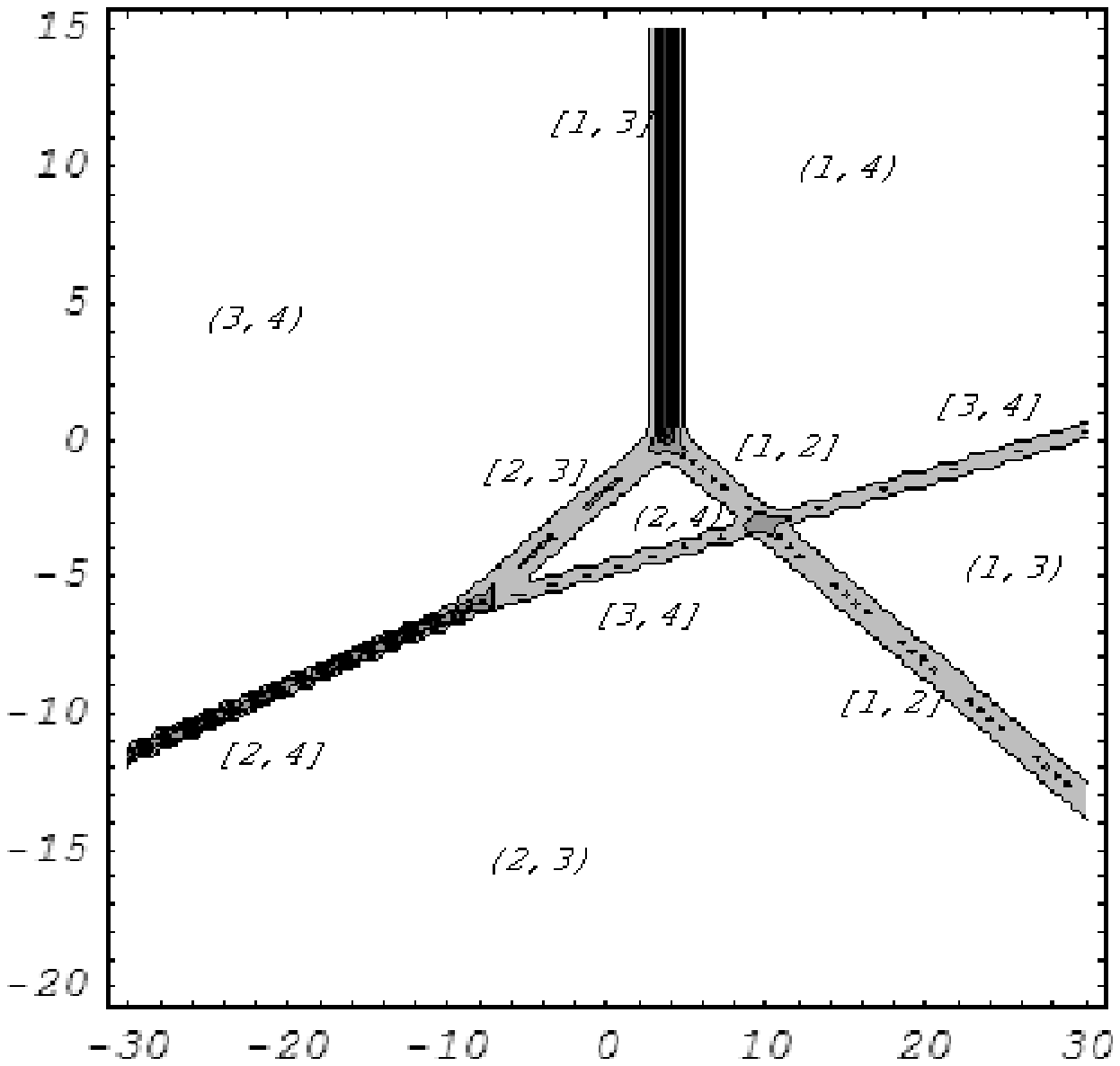}
\includegraphics[width=40mm,clip]{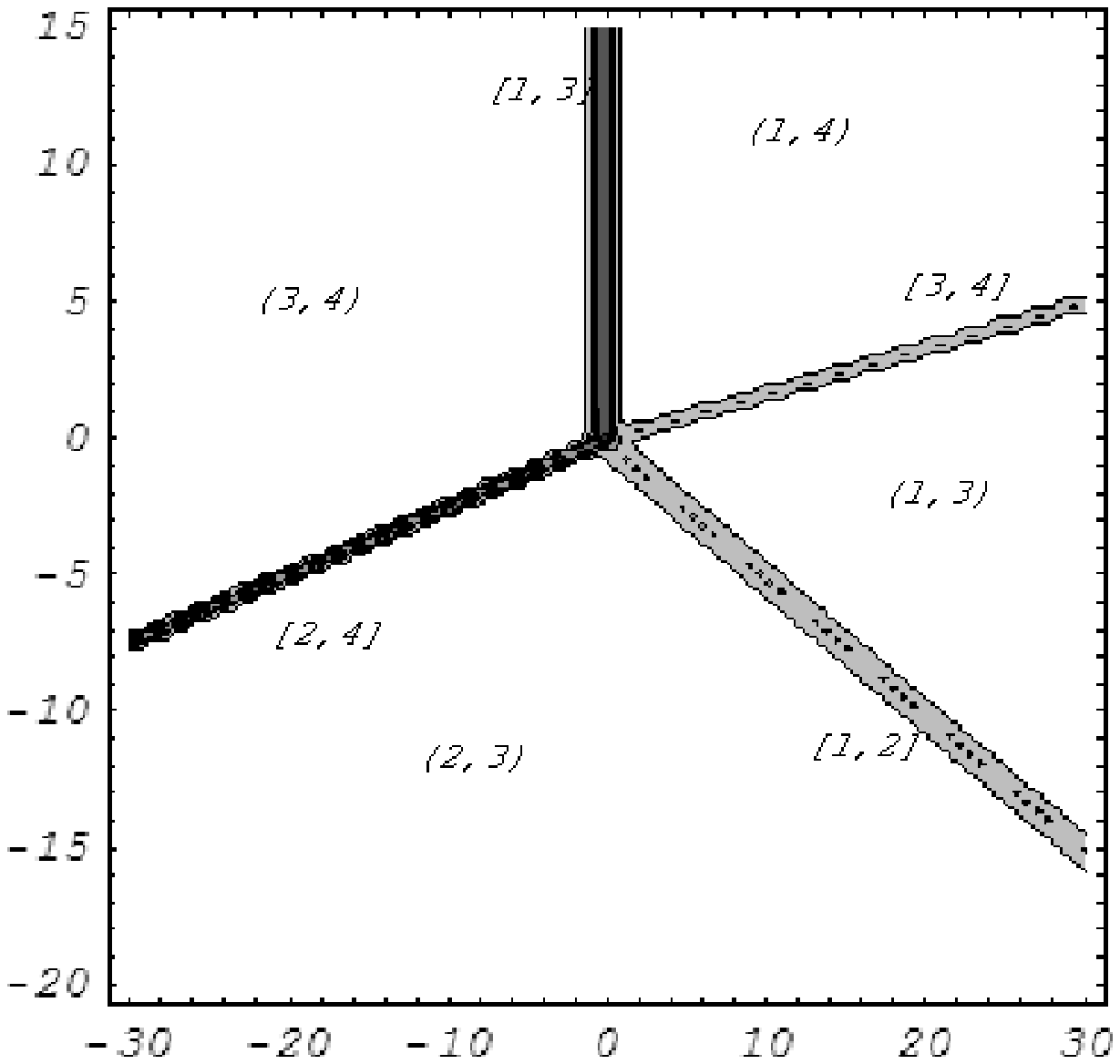}
\includegraphics[width=40mm,clip]{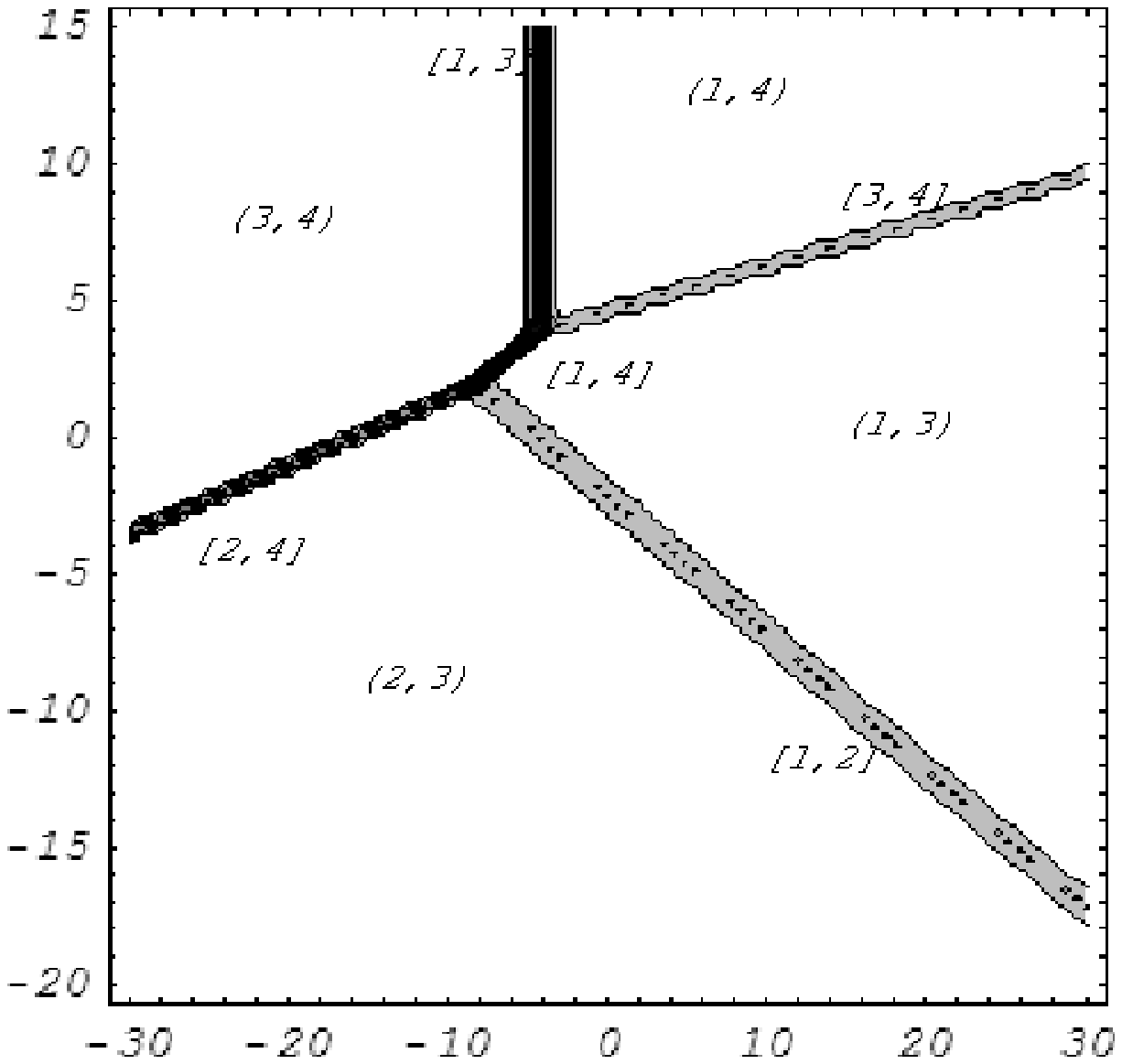}
\end{center}
\caption{Contour plots of the inelastic 2-soliton solution 
shown in Fig.~\ref{f:inelasticIs}. 
The dominant phase combinations $(m_1,m_2)$ and the index pairs 
$[i,j]$ that uniquely identify the line solitons are labeled.
Left: $t=-1$; center: $t=0$; right: $t=1$.}
\label{f:inelasticIc}
\end{figure}

\paragraph{Type I.} 
In this case $C_{1,2}=0$. 
The incoming solitons are [1,2] and [2,4], 
the outgoing solitons are [1,3] and [3,4]. 
The dominant phase combination as $x\to - \infty$ is $(3,4)$, 
while that as $x\to \infty$ is $(1,3)$. 
The interaction pattern is a combination of two Y-shape resonances,
as shown in Figs.~\ref{f:inelasticIs} and~\ref{f:inelasticIc}.
In particular, Fig.~\ref{f:inelasticIc} shows contour plots 
of the solution as well as the index pairs corresponding to each
of the intermediate and asymptotic solitons.
The interaction vertices, however, are not invariant:
For $t<0$ the three phases appearing at each interaction
vertex are respectively (2,3,4) and (1,2,3), corresponding respectively 
to solitons [2,3], [3,4], [2,4] and [1,2], [2,3], [1,3].  
But at $t=0$ two resonant stems collide, 
and the arrangement of solitons changes thereafter,
with the resonant vertices being characterized by the phases 
(1,2,4) and (1,3,4) for $t>0$, 
corresponding respectively 
to solitons [1,2], [1,4], [2,4] and [1,3], [1,4], [3,4].  
(An additional X-shape vertex is produced for $t<0$ by 
the asymptotic solitons [1,2] and [3,4]. This interaction is 
locally the ordinary 2-soliton interaction and the 
local maximum is less than $\frac12 (a_{1,2}+a_{3,4})^2 < 
\frac12 a_{1,4}^2$. We can also prove $u(0,0,0) < \frac12 
a_{1,4}^2$ similarly to the case of resonant 2-soliton 
interactions.) 

The rearrangement in the soliton configuration that happens at $t=0$
corresponds to the generation of a large-amplitude wave for $t>0$. 
The interaction arm is the intermediate soliton [2,3] for $t<0$ and 
[1,4] for $t>0$.
The first of these is always shorter than the asymptotic solitons
[1,3] and [2,4].
The second one, however, is taller than any of the others.
Moreover, the height of the interaction arm [1,4] is
$U_\mathrm{max}= \half a_{1,4}^2= \half (a_{1,j}+a_{j,4})^2>\half (a_{1,j}^2+a_{j,4}^2)\,$,
for $j=2,3$.
Thus, the height of the soliton [1,4] is always greater than 
the sum of the heights of the incoming and the outgoing solitons.  
As before, the maximum value of the interaction height relative to 
the height of the asymptotic solitons occurs in the case of equal
amplitudes: when $a_{1,2}=a_{2,4}=:a$, it is $U_\mathrm{max}= 2a^2$,
yielding a ratio of four to one. 
However, it should be noted that 
since $a_{1,3} > a$, $U_{\rm max}$ is 
less than four times the height of [1,3]-soliton. 
That is, the maximum height $U_{\rm max}$ is 
less than four times the height of the highest 
asymptotic soliton.  
If we impose still further the condition $a_{1,3}=a_{3,4}=a$, 
the solution degenerates to a Y-shape resonant solution.    

\begin{figure}[t!]
\kern-1.5\bigskipamount
\begin{center}
\includegraphics[width=42.5mm,clip]{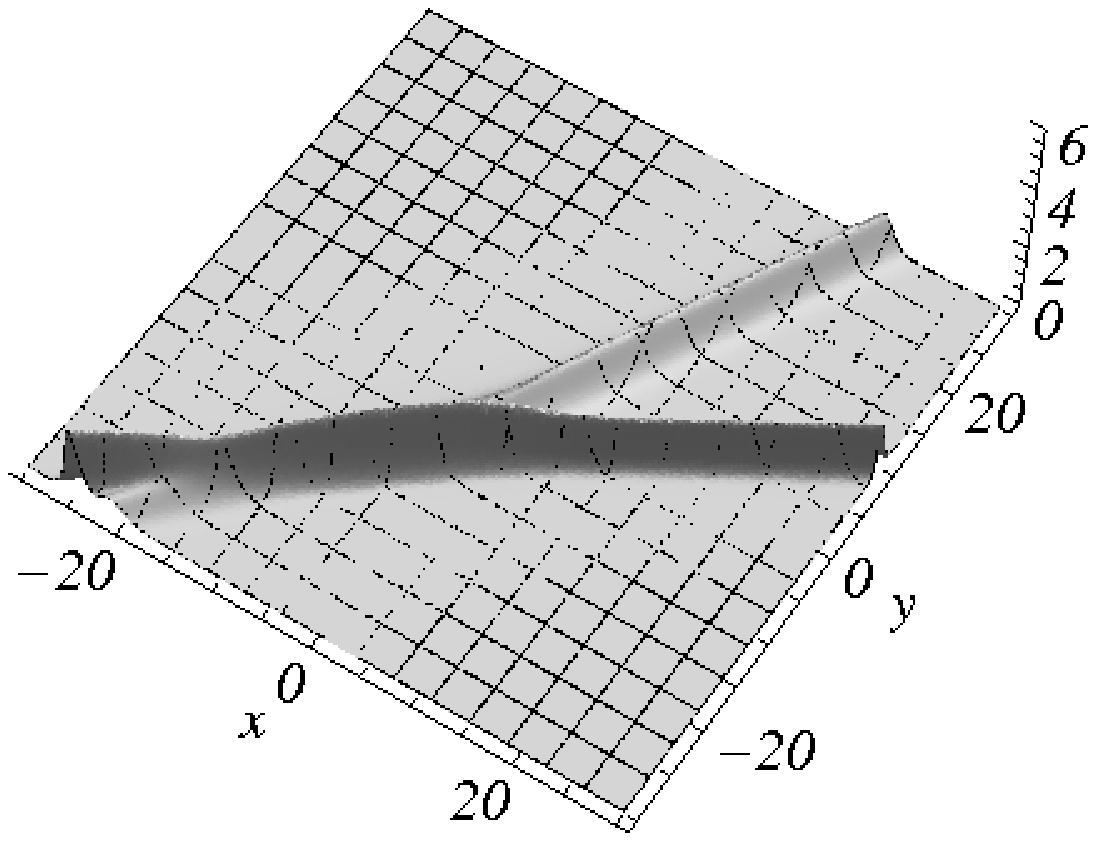}%
\includegraphics[width=42.5mm,clip]{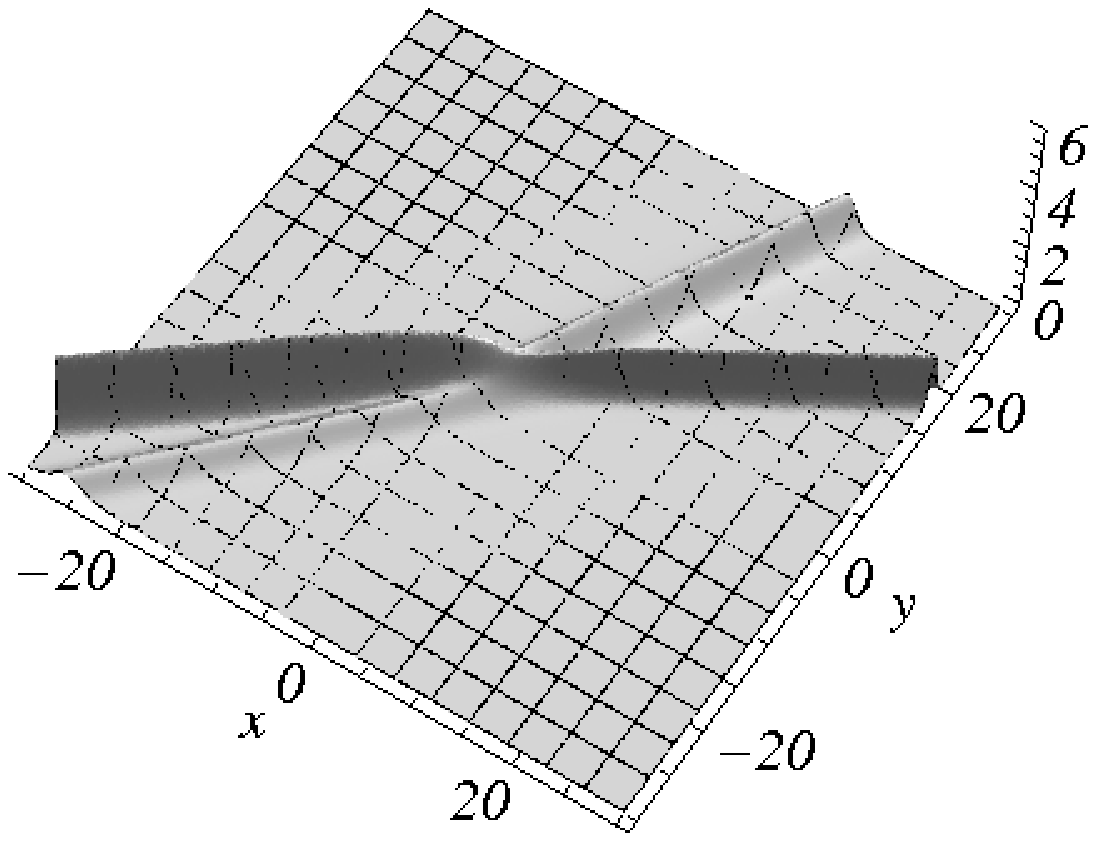}%
\includegraphics[width=42.5mm,clip]{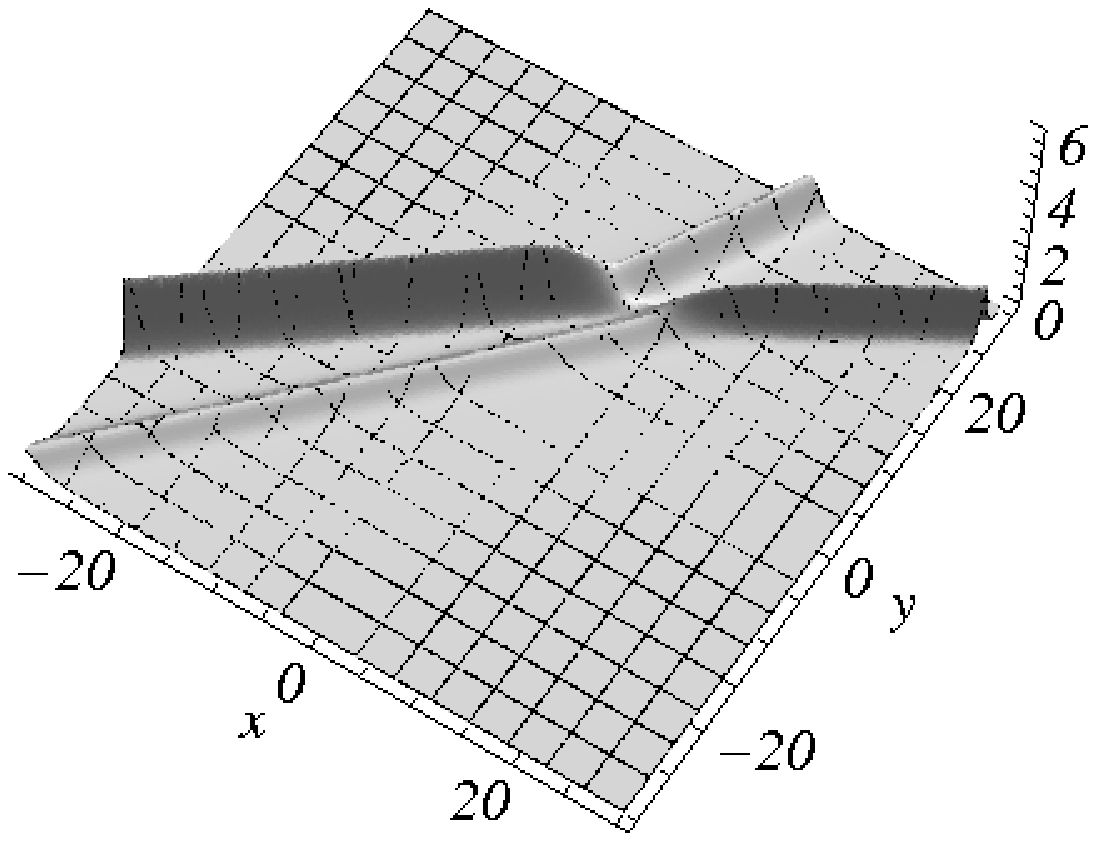}%
\end{center}
\smallskip
\caption{Time evolution of the inelastic 2-soliton solution obtained from
$C_\mathrm{III}$ in \eref{e:Ainelastic2s} with 
$(k_1,\dots,k_4)=(-0.5,-0.1,1,1.7)$.
Left: $t=-5$; center: $t=0$; right: $t=5$.}
\label{f:inelasticIIIs}
\medskip
\begin{center}
\includegraphics[width=40mm,clip]{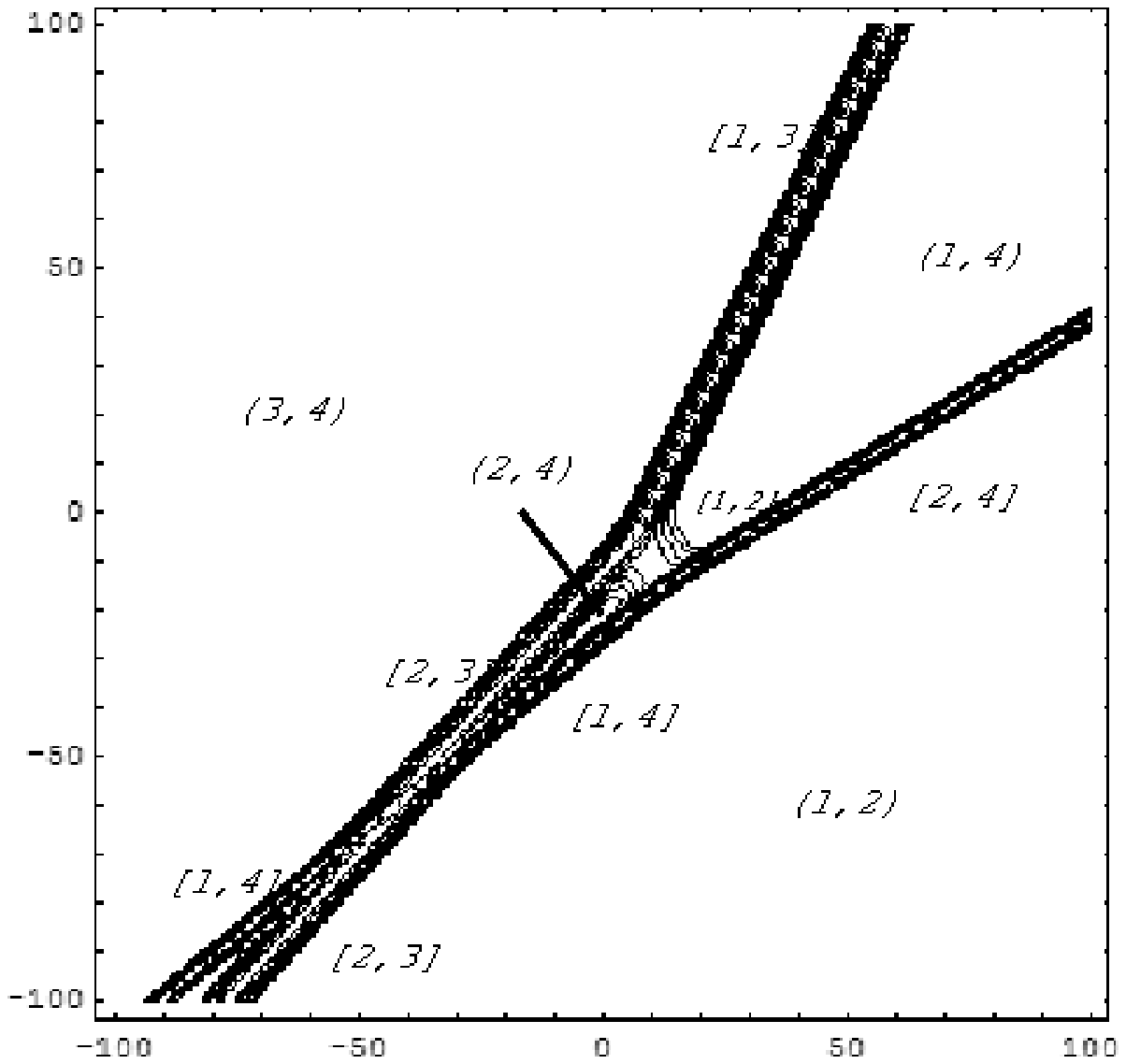}
\includegraphics[width=40mm,clip]{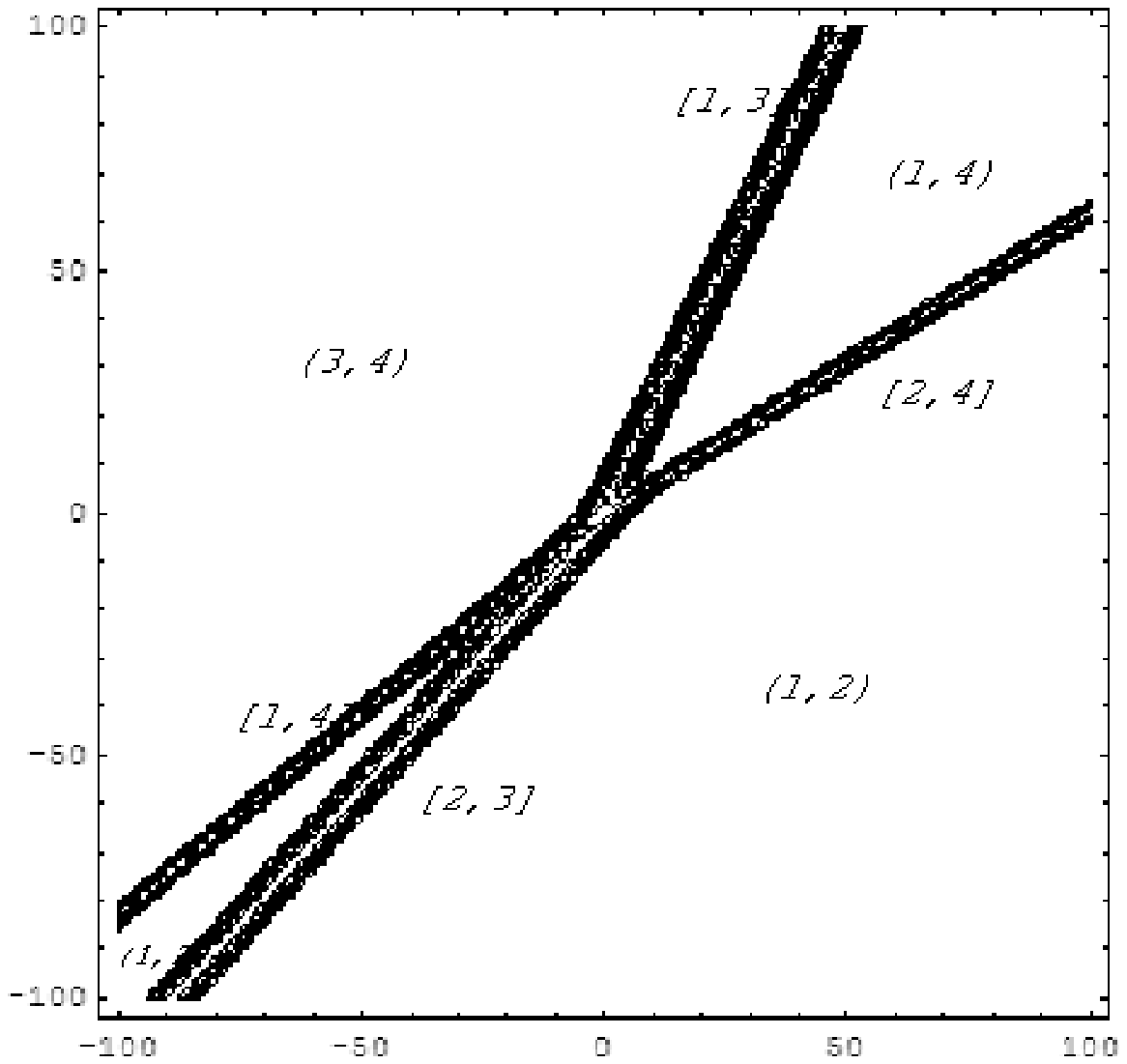}
\includegraphics[width=40mm,clip]{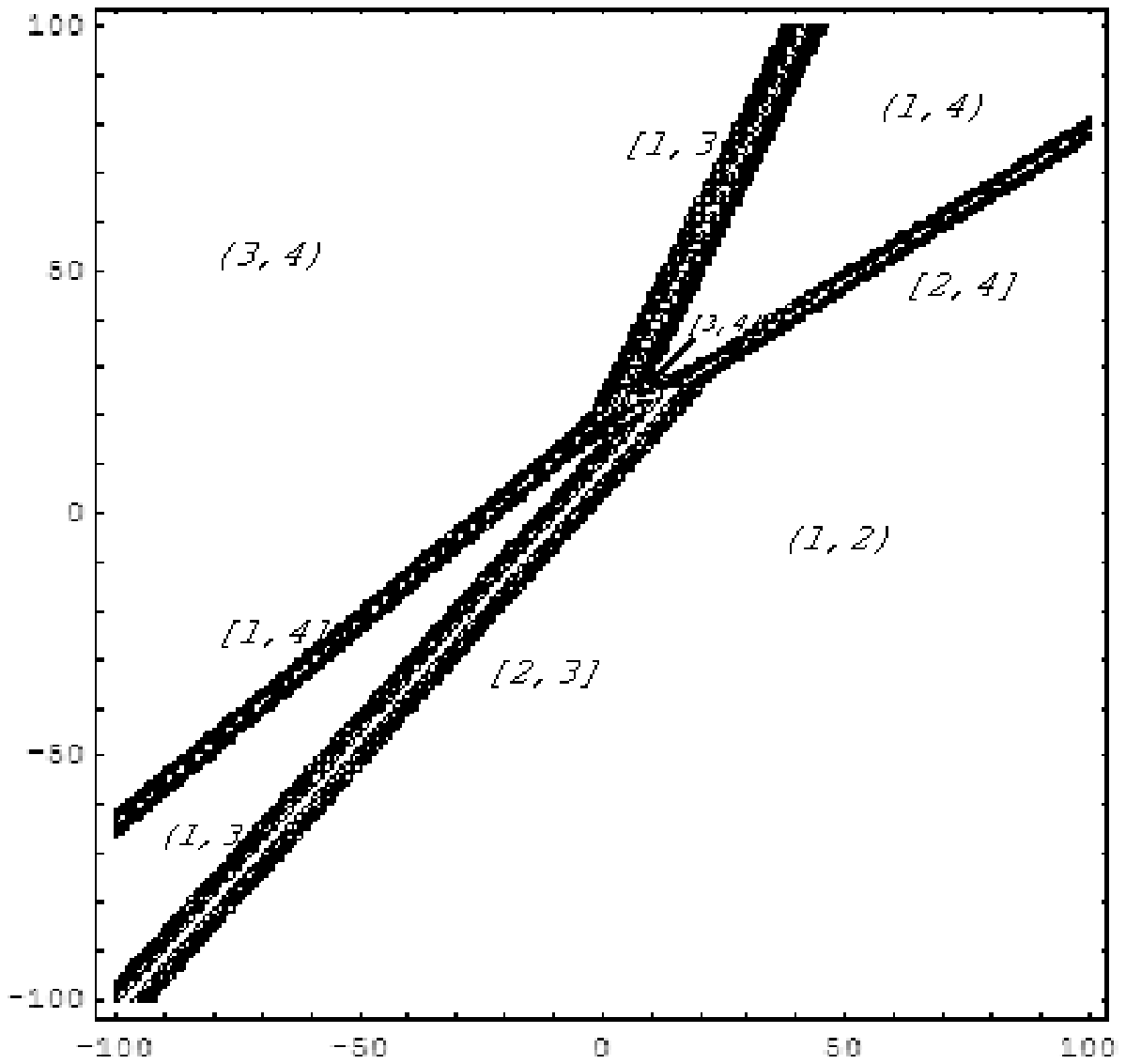}
\end{center}
\smallskip
\caption{Contour plots of the time evolution of the inelastic 
2-soliton solution shown in Fig.~\ref{f:inelasticIIIs}. 
Left: $t=-13$; center: $t=0$; right: $t=10$.}
\label{f:inelasticIIIc}
\end{figure}

\paragraph{Type II.}
In this case $C_{3,4}=0$.
The incoming solitons are [1,3] and [3,4],
the outgoing solitons are [1,2] and [2,4].
The whole solution is a time reversal of the inelastic 2-soliton 
solution of type~I: that is,
$u_{\rm II}(x,y,t)= u_{\rm I}(-x,-y,-t)$.
The dominant phase combination as $x\to - \infty$ and $x\to \infty$ 
are now respectively $(2,4)$ and $(1,2)$. 
The tall interaction arm is still the soliton [1,4], but
now it appears for 
$t<0$.

\paragraph{Type III.}
In this case $C_{2,3}=0$.
The incoming solitons are [1,4] and [2,3], 
the outgoing solitons are [1,3] and [2,4]. 
The dominant phase combination as $x \to -\infty$ is $(3,4)$, while that as $x \to \infty$ is $(1,2)$. 
The interaction dynamics is similar to that of solutions of type I,
as shown in Figs.~\ref{f:inelasticIIIs} and~\ref{f:inelasticIIIc}. 
The interaction pattern is a combination of two Y-shape 
resonances. The pattern differs depending on the magnitude 
relation between $k_1+k_4$ and $k_2+k_3$. 
Figures \ref{f:inelasticIIIs} and~\ref{f:inelasticIIIc} show 
the case $k_1+k_4 > k_2+k_3$. In this case,  
for $t < 0$, two Y-shape resonances consist of 
solitons [2,3], [1,3],[1,2] and solitons [1,4],[2,4],[1,2] 
and for $t > 0$, solitons [1,4],[1,3],[3,4] and 
solitons [2,3],[2,4],[3,4]. 
For $t < 0$, [1,4] and [2,3] solitons make an asymmetric 
interaction locally. This interaction generates only the 
height less than the height $U_{1,4}$ of the highest 
asymptotic soliton. At $t=0$, the exchange of combination 
of two Y-shape resonances takes place and four solitons 
interact near the origin of the $xy$-plane. We can again prove 
$u(0,0,0) < U_{1,4}$ as in the resonance 2-soliton interactions. 
Accordingly, $U_{1,4}$ is the maximum height in this interaction. 
For $k_1+k_4 < k_2+k_3$, the combinations of Y-shape resonances 
change and an asymmetric interaction of [1,4] and [2,3] solitons 
takes place for $t > 0$. However, the result for 
the maximum height is the same before.
In the case $k_1+k_4 = k_2+k_3$, [1,4] and [2,3] solitons are 
parallel and make an over-taking interaction. At $t=0$, these 
two soliton may coalesce into one peak, the height of which is 
$U_{1,4}-U_{2,3}$ as described in the subsection on 
asymmetric 2-soliton interactions, and at the same time 
the exchange of combination of two Y-shape resonances takes place. 
So, in this case the maximum height is also $U_{1,4}$.  

\paragraph{Type IV.} 
In this case $C_{1,4}=0$. 
The incoming solitons are [1,3] and [2,4], 
the outgoing solitons are [1,4] and [2,3]. 
Such a solution 
is a time-reversal version of the inelastic solution of type~III:
that is, $u_{\rm IV}(x,y,t)= u_{\rm III}(-x,-y,-t)$.

\smallskip
\subsection{Generation of large-amplitude waves}

\begin{figure}[t!]
\kern-\bigskipamount
\begin{center}
\kern-0.2em
\includegraphics[width=44mm,clip]{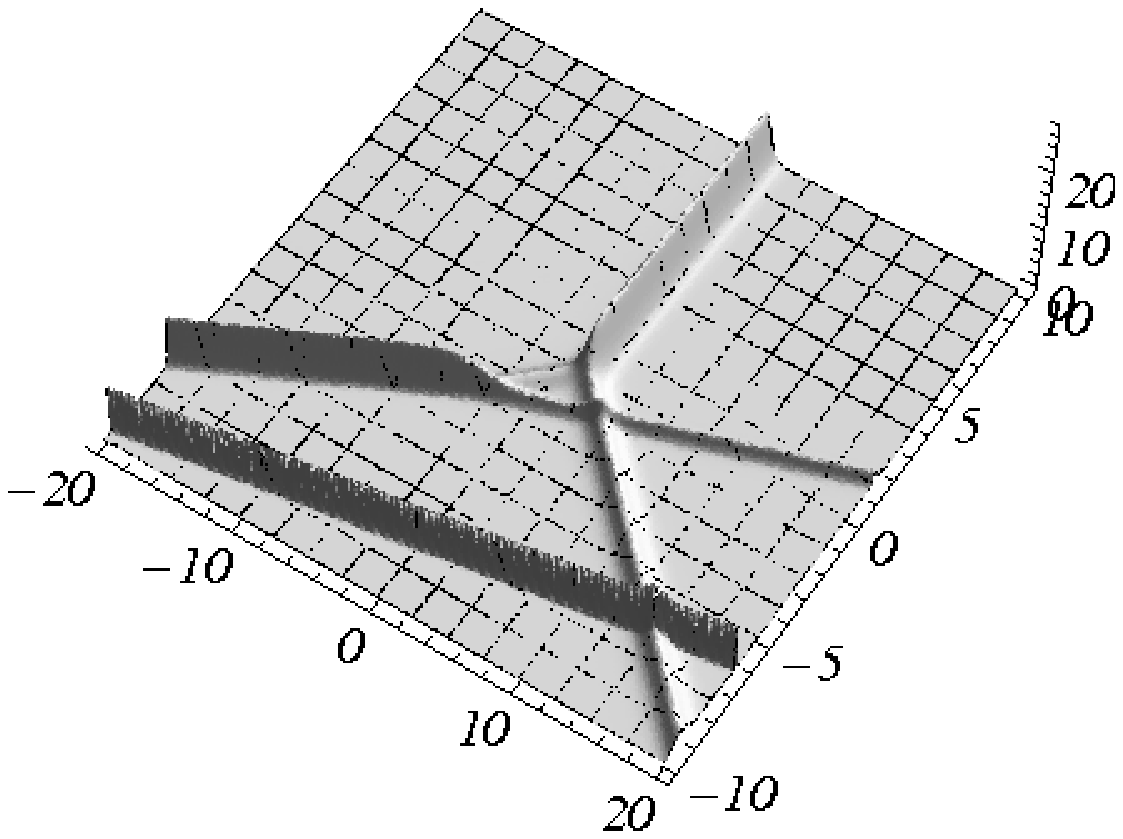}%
\includegraphics[width=44mm,clip]{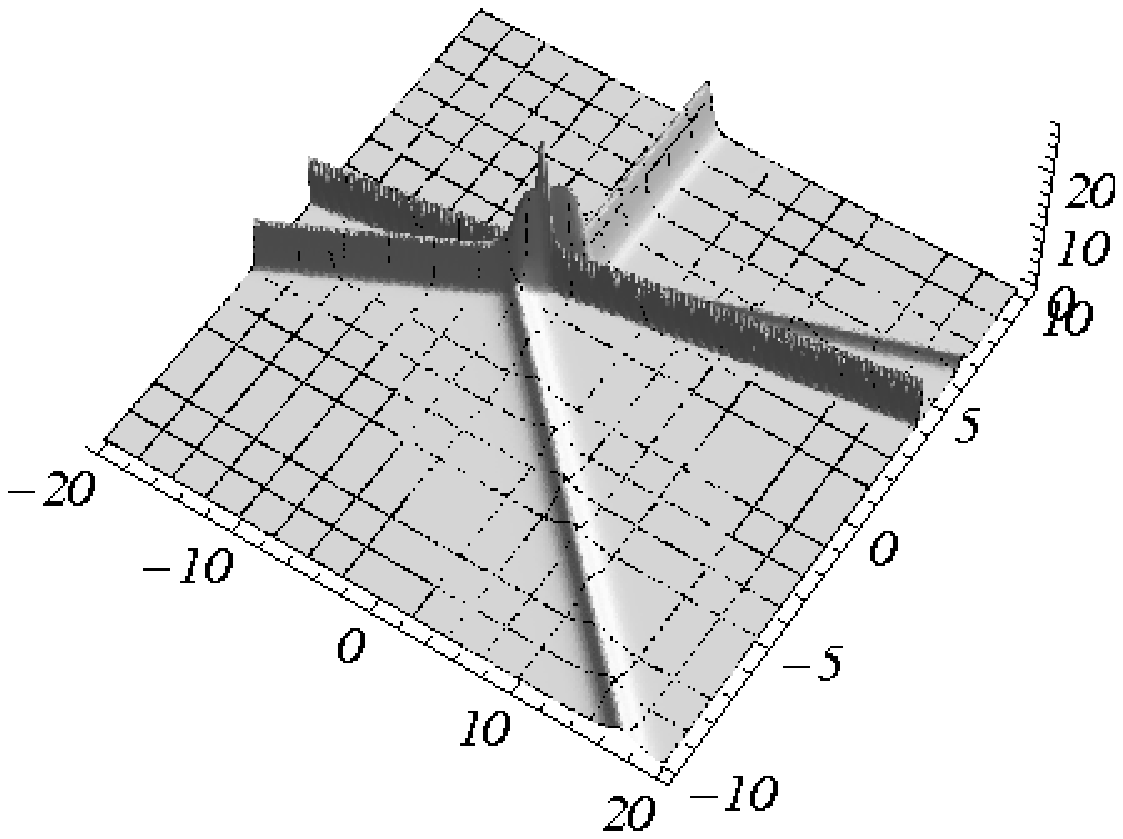}%
\includegraphics[width=44mm,clip]{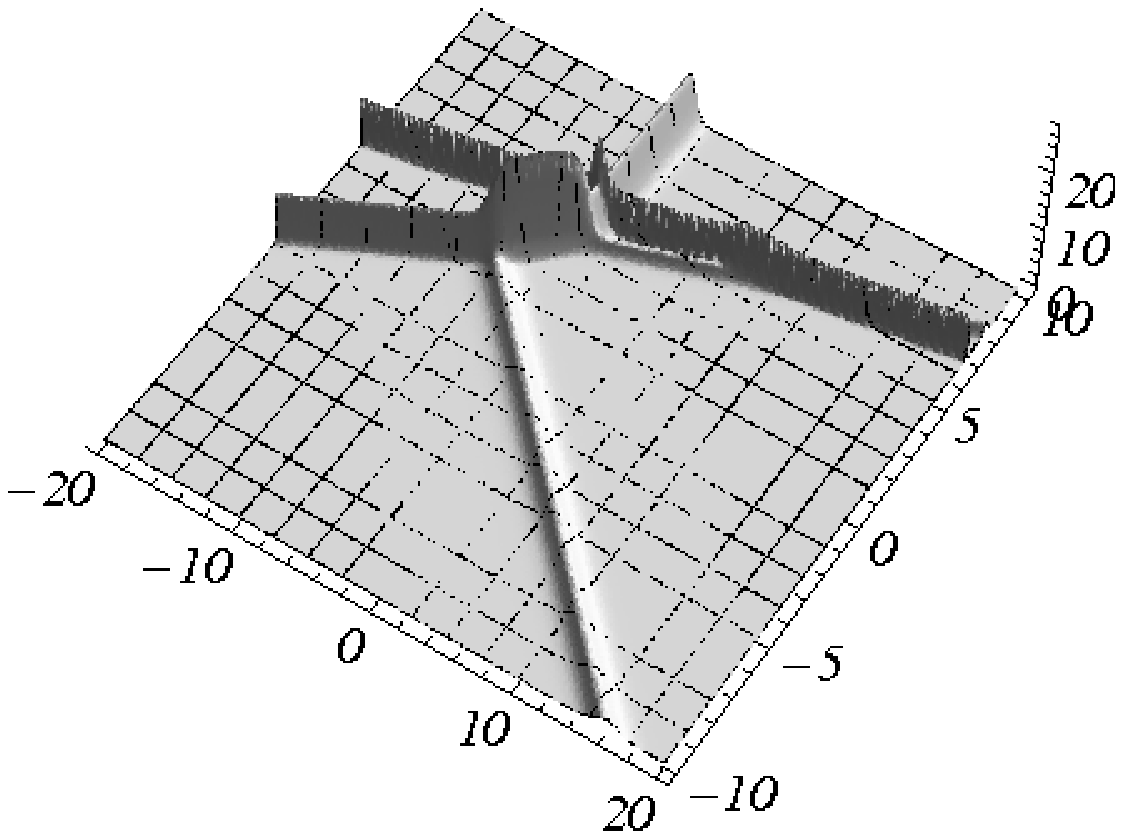}%
\end{center}
\medskip
\begin{minipage}[b!]{0.64\textwidth}
\caption{A solution of KP generating an extreme wave,
as produced by the coefficient matrix in \eref{e:bigwave}
with $(k_1,\dots,k_6)= (-2,0,2,4,
4.01,8)$,
$\theta_{1;0}=\dots\theta_{5;0}=0$
and $\theta_{6;0}= 200$.
Above, left: $t=-0.4$; above, center: $t=0.7$; above, right: $t=1$.
The plot to the right shows the maximum wave height 
as a function of time.}
\label{f:bigwave}
\end{minipage}\kern1.4em 
\begin{minipage}[b!]{0.35\textwidth}
\includegraphics[width=38.5mm,clip]{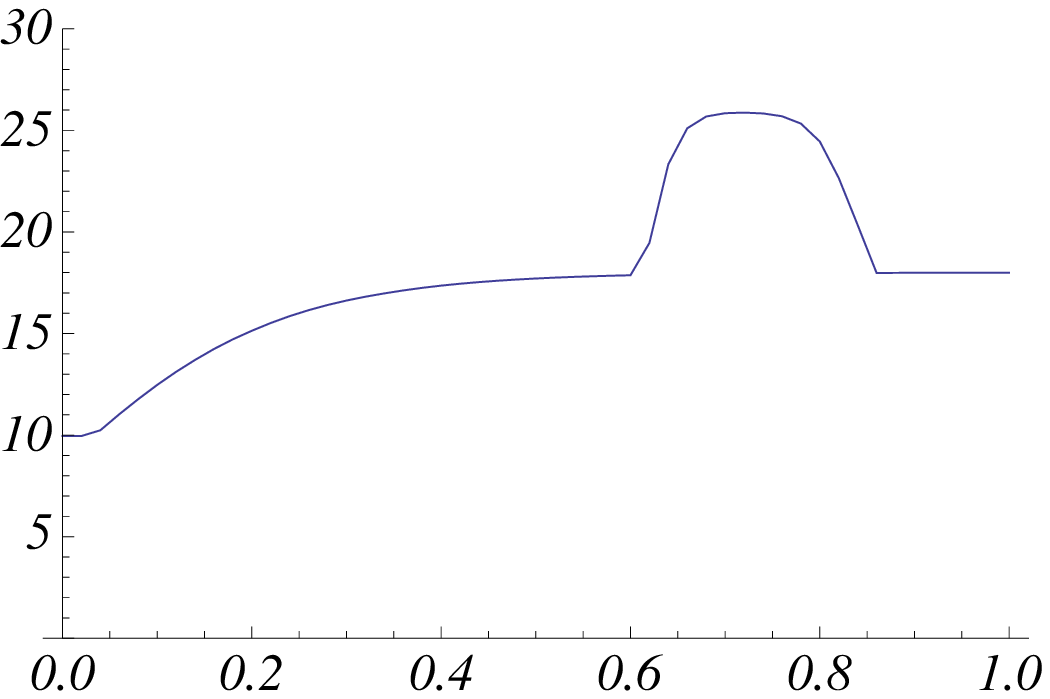}\hglue-6em%
\end{minipage}%
\smallskip
\end{figure}

None of the three types of elastic 2-soliton solutions describes 
the generation 
of large-amplitude waves from the interaction of lower-amplitude ones,
since the interaction pattern of ordinary and asymmetric solutions
is stationary, and the intermediate soliton [1,4] in resonant solutions 
is present at all times.
On the other hand, inelastic solutions of type~I do have the effect of an amplification of the maximum wave height. 
Moreover, when this kind of solution is embedded 
in a larger soliton complex,
further increases of the wave height may result from the interaction
of the interaction arm with the other solitons in the complex.
As an example, Fig.~\ref{f:bigwave} shows the large-amplitude wave
produced by the interaction generated by the $3\times6$ coefficient matrix
\begin{equation}
C= \begin{pmatrix} 1 &1 &1 &0 &0 &0\\ 0 &0 &1 &1 &0 &0\\ 0 &0 &0 &0 &1 &1
  \end{pmatrix}\,.
\label{e:bigwave}
\end{equation}
As evident from Fig.~\ref{f:bigwave}, the interaction among the solitons results
in the temporary generation of an extreme wave whose height 
exceeds four times that of the highest asymptotic soliton. 

This discussion suggests the following physical mechanism of generation of extreme waves: 
(i) several solitons are generated by external sources; 
(ii) two of those solitons generate a large-height interaction arm, 
as in the case of ordinary 2-soliton solutions, or that of 
inelastic solutions of type~I;
(iii) this interaction arm interacts with one of the 
other solitons or interaction arms, in which case the wave height 
can be many times higher that of each asymptotic soliton;
(iv) after the interaction, the wave amplitude decreases.  

\subsection{A method to predict maximum interaction amplitudes from asymptotic data}
\label{s:maxamp}

The above results yield a method to predict the possible 
maximum amplitude of a soliton interaction based only on information 
about the asymptotic line solitons.
Consider first the case in which there are two line solitons
as $y \to \pm\infty$. 
Let $(a_1^-,d_1^-)$ and $(a_2^-,d_2^-)$ be the amplitudes and directions
of the incoming line solitons, and 
$(a_1^+,d_1^+)$ and $(a_2^+,d_2^+)$ be those of the outgoing line solitons,
both sorted in order of increasing values of~$d$.
If $d_1^-= d_1^+$ or $d_2^-= d_2^+$, 
the incoming and outgoing line solitons coincide, 
implying that one has an elastic soliton interaction. 
Otherwise one has an inelastic soliton interaction.  
For an elastic soliton interaction,
the interaction amplitude is computed as follows.
Let $\kappa_{n,\pm} = (d_n \pm a_n )/2$ for $n=1,2$.
[The superscript $\pm$ in the soliton parameters $a_n$ and $d_n$ 
can of course be dropped for elastic solutions.]
Then define the phase parameters $k_1,\dots,k_4$ to be the values
$\kappa_{1,\pm},\kappa_{2,\pm}$ rearranged in order of increasing size,
such that $k_1<k_2<k_3<k_4$.
In this way each soliton is uniquely identified by the index pair 
$[i_n,j_n]$ that labels the position of $\kappa_{n,-}$ and $\kappa_{n,+}$
(respectively) in the list $k_1,\dots,k_4$.
The type of index overlap determines the type of soliton interaction
\cite{prl2007,jpa2008,jpa2004}, 
and the interaction amplitude is then obtained from the calculations
described earlier.
A similar method applies in the case of an inelastic interaction.  
In this case, however, one needs the asymptotic data both as
$y\to-\infty$ and as $y\to\infty$ in order to uniquely identify
the type of interaction.

If the number of solitons is greater than two, one selects 
any two neighboring solitons and perform the above procedure
to compute the possible maximum amplitude. 
One then repeats these steps for all possible pairwise 
combinations of solitons. 
Note, however, that, unlike the case of two solitons, 
this procedure does not yield a precise estimate, for two reasons:
(i) whether or not the theoretical maximum amplitude in each pairwise
interaction is realized depends on the details of the
soliton configuration;
(ii) these high-amplitude intermediate solitons resulting from pairwise 
interactions can in some cases interact among themselves producing 
solitons of even higher amplitude.
Again, whether or not this happens depends on the details of the 
soliton configuration.
A true upper bound can be obtained: 
$U_\mathrm{max}= \half(k_\mathrm{max}-k_\mathrm{min})^2$.
This theoretical maximum, however, is realized only 
in a small number of soliton interactions, as should already be
evident from the case of 2-soliton solutions.

\section{Numerical simulations}

We now describe numerical simulations of multi-soliton interactions 
of the KPII equation.
The numerical simulation of multi-soliton solutions is
particularly important,
since at present no analytical methods exist to investigate the 
stability of such solutions using either the inverse scattering transform 
or other techniques.

\subsection{A computational method for line-soliton solutions}

The numerical integration of the KP equation poses a number of challenges
(e.g., see \cite{Klein} and references therein).
In particular, when simulating soliton solutions 
one must take into account that line solitons are not localized objects, 
but they extend through the boundaries of any finite computational window.
The approach we used here is based on the one in 
\cite{tsuji1,oikawa1,oikawa2,tsuji2,Wineberg},
but with different boundary conditions.
For the $x$-direction, we set our computational window to be wide enough
that any initial solitary waves are far away from boundary. 
This allows us to use periodic boundary conditions
and to compute $x$-derivatives with spectral methods.
For the $y$-direction we employ the windowing method \cite{Schlatter},
which has its roots in signal processing, and 
where the windowing operation allows the spectral analysis of non-periodic
signals. 
We use the following window function:
$W(y)=10^{-a^n|2y/L-1|^n}\,$,
where $L$ is the length of the computational window in the $y$-direction, 
and $a$ and $n$ are parameters. 
Here we set $a=1.111$ and $n=27$. 
We then transform the solution as follows:
\begin{equation}
q(x,y,t)= W(y)\,u(x,y,t)\,,
\qquad Q(x,y,t)= W(y)\,u^2(x,y,t)\,.
\label{e:uqdef}
\end{equation}
Substituting this into the KPII equation, we obtain 
\begin{equation}
\big(-4q_t+ 3Q_x+ q_{xxx}\big)_x
 + 3q_{yy} - 6W_yu_y + 3W_{yy}u = 0\,.
\label{e:qpde}
\end{equation}
All terms in this equation vanish at the boundaries in the $y$-direction. 
This makes it possible to apply pseudospectral methods to compute 
$y$-derivatives.
We then integrate \eref{e:qpde} in time in the Fourier domain using 
Crank-Nicholson differencing and an iterative method.
Once $q(x,y,t)$ is obtained, 
$u(x,y,t)$ is recovered from~\eref{e:uqdef}.
But note that the formula for $u(x,y,t)$ becomes ill-conditioned
near the boundaries in the $y$-direction, where $W(y)$ tends to zero.
Near these boundaries,
we thus correct the solution using information about the 
soliton behavior.
All the simulations were performed on a grid with $8192\times1024$ points, 
$\Delta x = \Delta y= 0.1$ and $\Delta t= 0.005$.
Figures \ref{web-num}--\ref{inelastic-num-soliton} below 
show the resulting field $q(x,y,t)$.  Note that, to make the interactions 
more evident, only a small portion of the computational domain
is often shown.

\begin{figure}[t!]
\begin{center}
\includegraphics[width=68mm,clip]{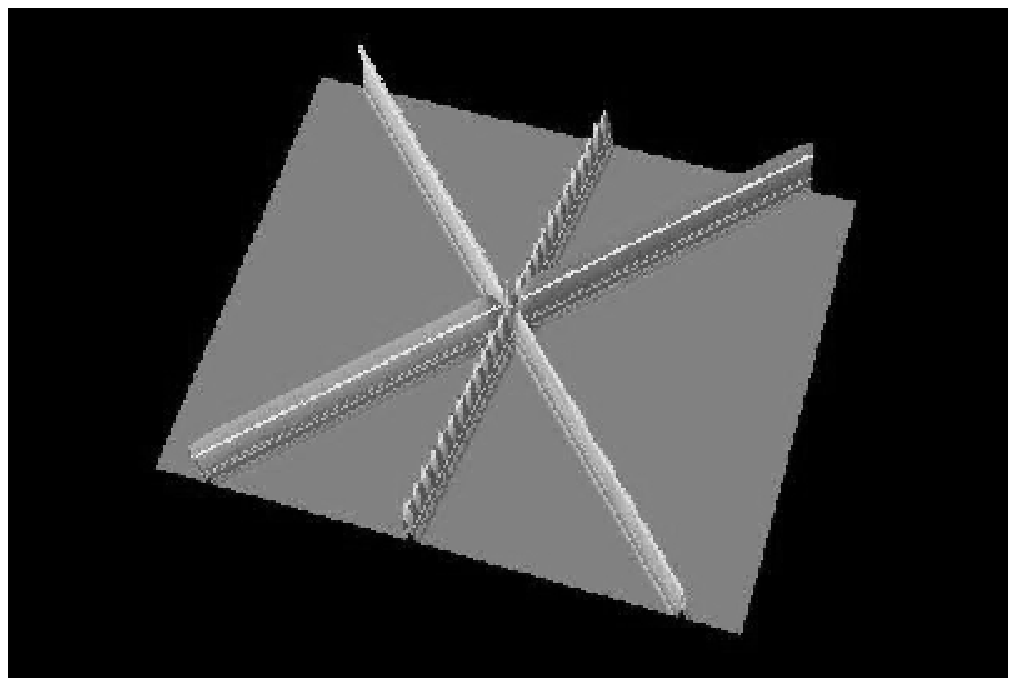}\kern-2em
\includegraphics[width=68mm,clip]{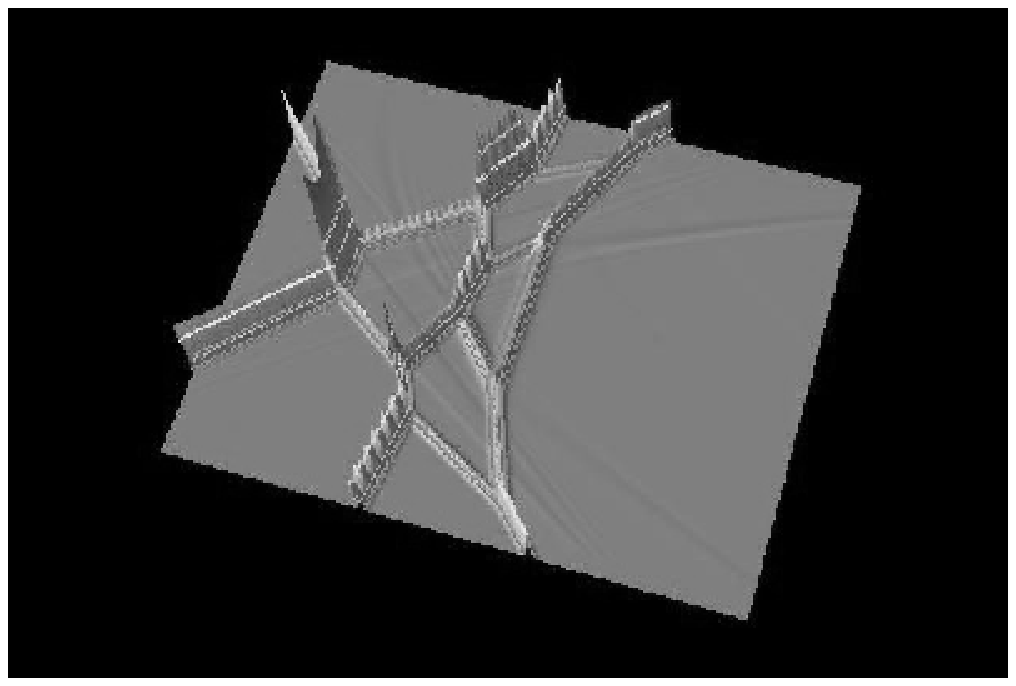}%
\end{center}
\caption{Numerical time evolution of the fully resonant 3-soliton interaction;
$(k_1,\dots,k_6)=(-2.5,-1.5,-0.5,0.5,1.5,2.5)$.
Left: $t=0$; right: $t=10$.}
\label{web-num}
\vskip3\medskipamount
\begin{center}
\includegraphics[width=68mm,clip]{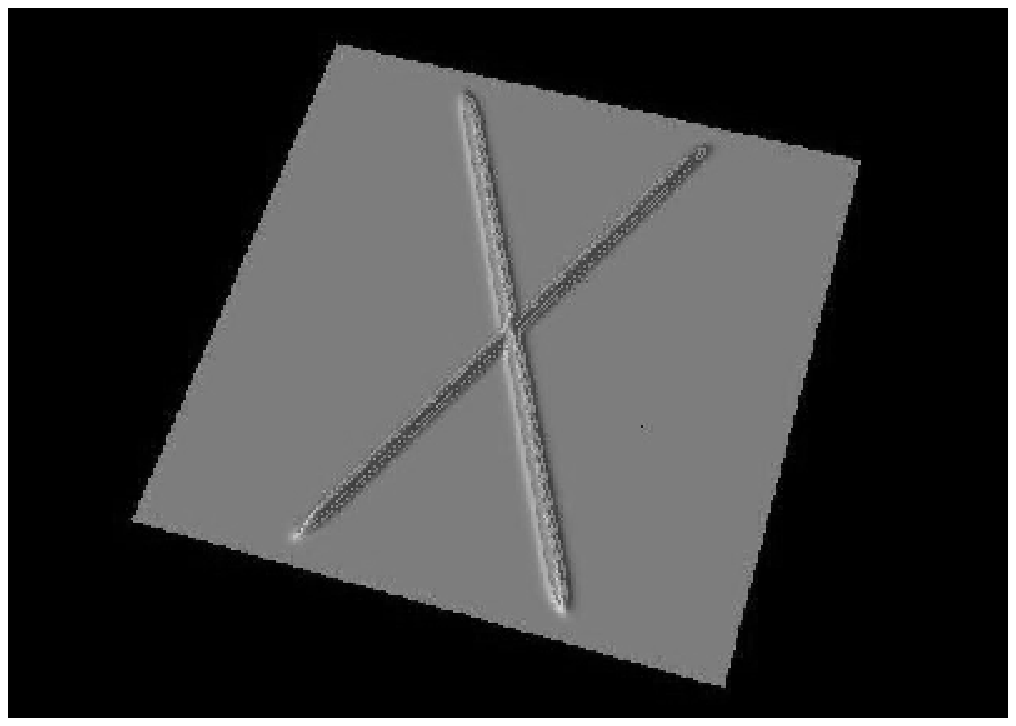}\kern-2em
\includegraphics[width=68mm,clip]{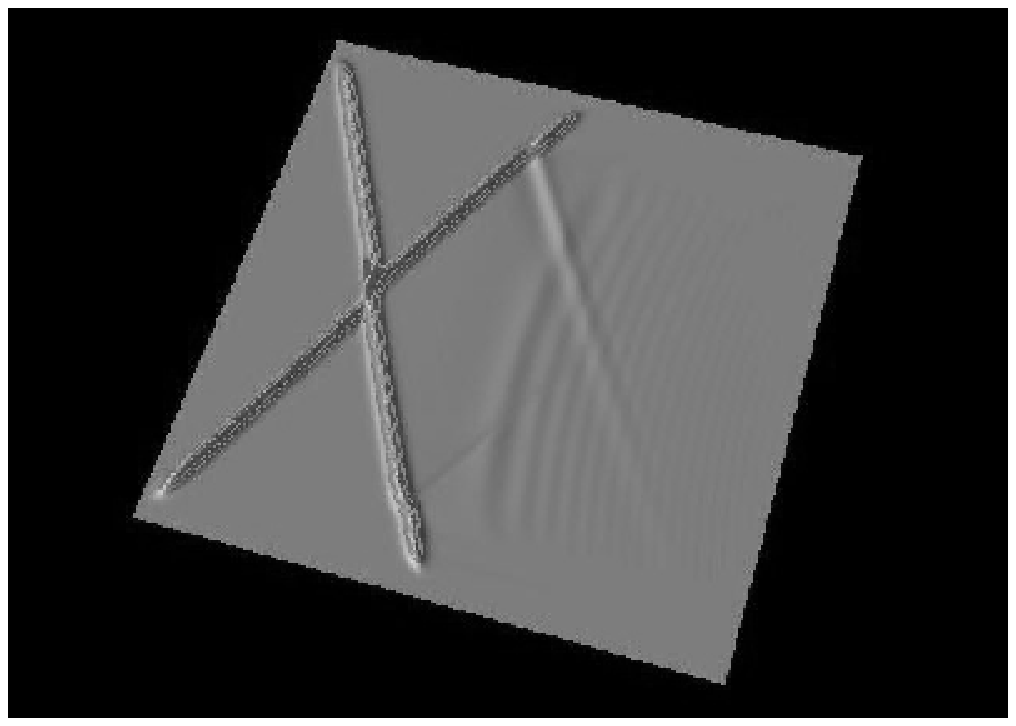}%
\end{center}
\caption{Numerical time evolution of a linear superposition of 
line solitons generating an ordinary 2-soliton interaction;
$(k_1,\dots,k_4)=(-1,-0.001,0,1)$.
Left: $t=0$; right: $t=50$.
The IC is not an exact 2-soliton solution, and some dispersive waves 
are generated, but the results suggest the stability of the 
soliton interactions.}
\label{o-type-num}
\vskip3\medskipamount
\begin{center}
\includegraphics[width=68mm,clip]{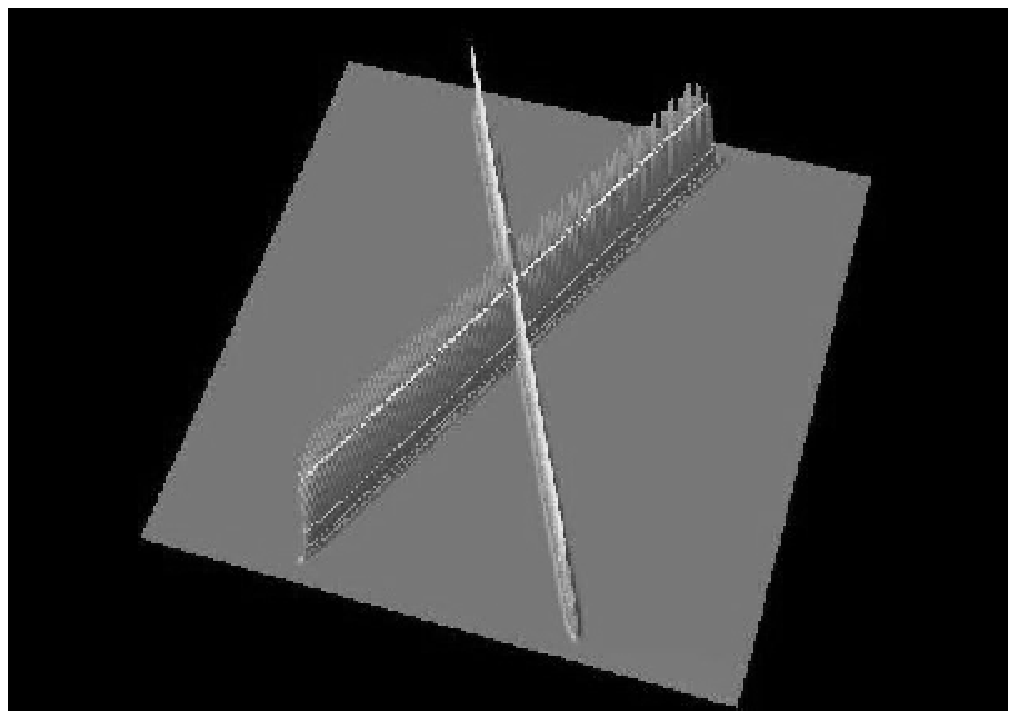}\kern-2em
\includegraphics[width=68mm,clip]{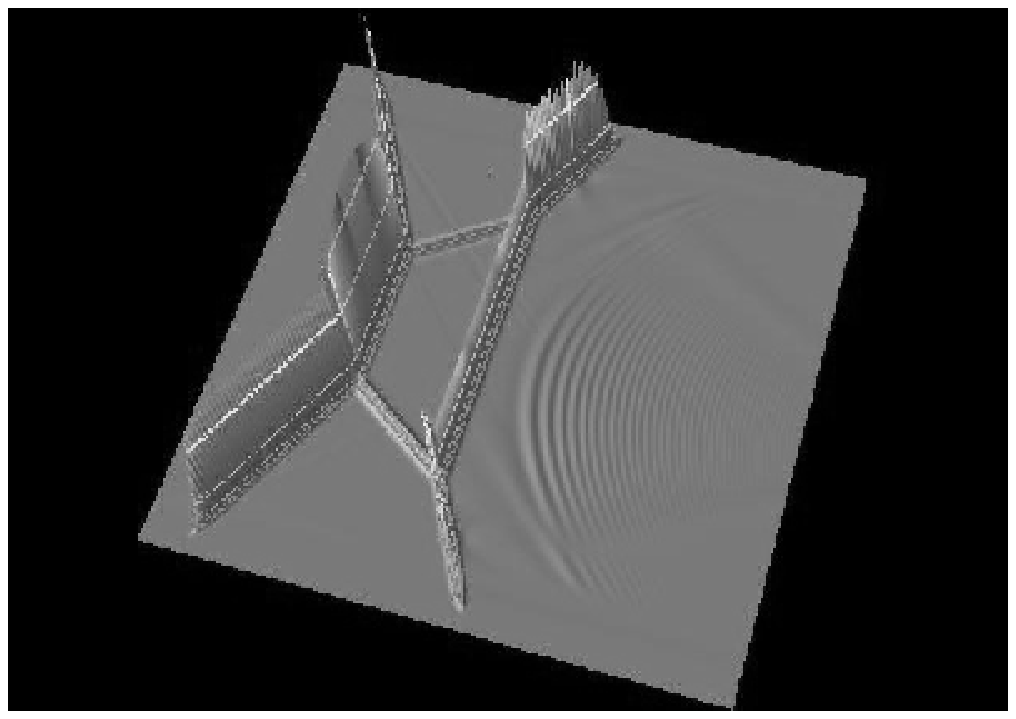}%
\end{center}
\caption{Numerical time evolution of a linear superposition of 
line solitons generating a resonant 2-soliton interaction;
$(k_1,\dots,k_4)=(-2,-1,1,2)$.
Left: $t=0$; right: $t=12.5$.
Again, dispersive waves are generated, but the interaction appears to be
stable.}
\label{t-type-num}
\kern-2\medskipamount
\end{figure}


\begin{figure}[b!]
\begin{center}
\includegraphics[width=\textwidth,clip]{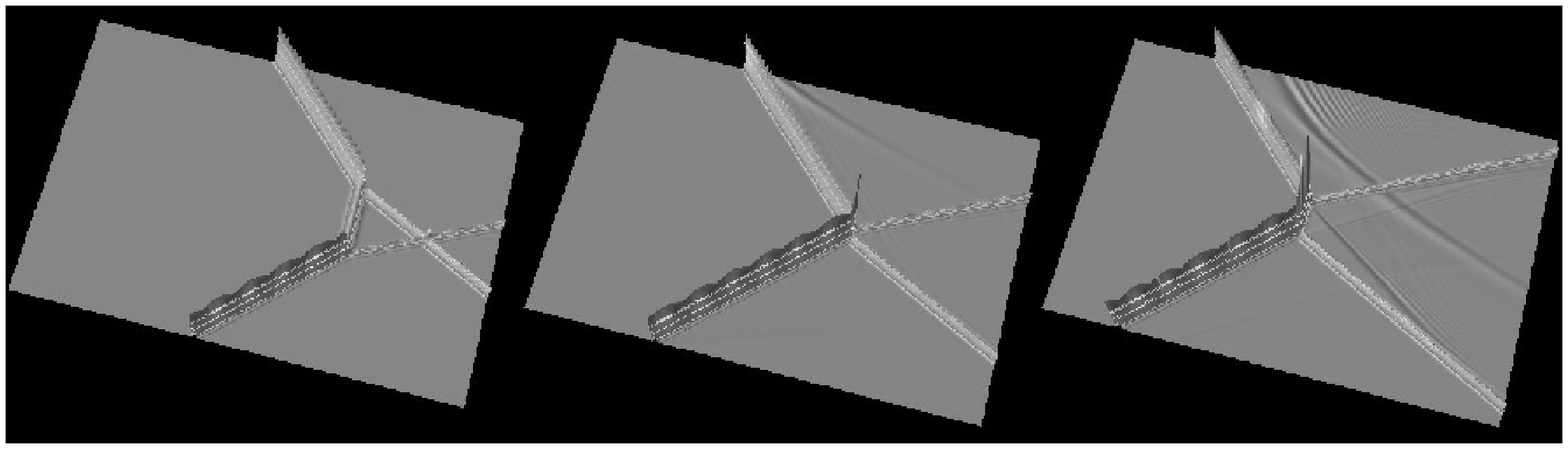}
\end{center}
\caption{Numerical time evolution of an inelastic 2-soliton interaction.
Left: $t=0$; center: $t=2.5$; right: $t=5$.}
\label{inelastic-num-soliton}
\kern-\medskipamount
\end{figure}

\subsection{Numerical simulations of line-soliton interactions}

Figure~\ref{web-num} shows the field $q(x,y,t)$ corresponding to the
numerical time evolution of an initial condition (IC)
consisting of an exact fully resonant 3-soliton solution. 
As evident from the figure,
the numerical solution accurately reproduces the web structure 
observed in the exact solution.  
This result confirms that the numerical method described above
can indeed effectively simulate multi-soliton 
interactions, and at the same time provides a first indication
that such solutions are stable.

Next we describe the time evolution of ICs
consisting of a linear superposition of two line solitons.
Figures~\ref{o-type-num} and \ref{t-type-num} 
show the cases where the amplitudes and directions were chosen so as
to correspond respectively to an ordinary and resonant interaction.
We emphasize that, in both cases, the initial state is not an 
exact two-soliton solution. 
Indeed, in both cases the numerical solution shows the presence of 
radiative component of small amplitude.
Nonetheless, the results do provide a further check of the
stability of 2-soliton interactions.
Note in particular that the characteristic ``box'' of the 
resonant solution is generated numerically.
Similar results were obtained by numerically computing the time 
evolution of an IC consisting of a linear superposition of two
line solitons corresponding to an asymmetric interaction and of
three line solitons corresponding to a fully resonant solution.
Finally, Fig.~\ref{inelastic-num-soliton} shows the time evolution
of an IC corresponding to an inelastic interaction.

Importantly, when the IC is not an exact 2-soliton solution,
the numerical solutions show that the height of the interaction arm 
tends monotonically in time to the value for the corresponding exact soliton
as obtained in section~\ref{s:interactions}
For example, in Fig.~\ref{o-type-num}
the height of interaction arm increases monotonically in time 
from its initial value of 2
[owing to an IC that is just a linear superposition 
of two line solitons],
approaching asymptotically the value corresponding to ordinary 
2-soliton interactions.
Conversely, for an asymmetric solution
the height of the interaction arm decreases in time, 
again approaching asymptotically the value for asymmetric 2-soliton 
interactions.
These results extend the validity and usefulness of the analysis 
in section~\ref{s:interactions}

The above results suggest that multi-soliton solutions of KPII 
are robust and stable,
and, moreover, that even when the solution contains 
non-solitonic components, 
only the information about the asymptotics line solitons contributes 
to determine the intermediate interactions of the line solitons.
We also suspect that, 
as the radiative components disperse,
asymptotically in time the solution will be closely approximated by 
an exact soliton solution,
similarly to the case of (1+1)-dimensional soliton equations.
Of course all of these conjectures must be carefully tested and
validated with extensive numerical simulations, which are beyond
the scope of this work.

\section{Conclusions}

We have studied the amplitude of soliton interactions of the KPII equation,
and we have discussed a possible mechanism for the generation of
large-amplitude waves.
Ordinary $N$-soliton interactions with $N\ge3$ can also briefly produce 
large amplitude waves if all the solitons intersect simultaneously 
at the same point in the $xy$-plane. 
Note however that the event of all solitons intersecting simultaneously 
at a single point can be considered to be statistically speaking unlikely
in a multi-soliton complex,
with a likelihood decreasing as the number of solitons increases. 
In this sense, therefore, the mechanism described in this paper,
involving inelastic 2-soliton solutions (possibly embedded in a
larger soliton complex, as in Fig.~\ref{f:bigwave})
represents the most likely way to generate large-amplitude waves. 

We also proposed a method to determine the maximum amplitude
resulting from the interaction of two line solitons. 
The calculation of the maximum amplitude 
is based on the framework of exact line-soliton solutions, 
but it may also be useful for solutions where a non-solitonic component 
is present,
if line-soliton solutions of KPII are indeed proven to be stable,
since then, even if one starts from an initial state that is not 
an exact soliton solution, the radiative portions of the solutions
will disperse away, and asymptotically in time one will approach 
a state consisting of an exact soliton solution.
An interesting open problem will be to develop an algorithm to compute the 
actual maximum amplitude generated by any multi-soliton configuration.
 
Finally, we implemented an algorithm to numerically integrate solutions of
the KPII equation containing multi-soliton complexes, 
and we discussed the results of numerical simulations.
These results show the robustness of all types of line-soliton solutions 
of KPII, including those exhibiting web-like structure. 
We also confirmed numerically that multi-soliton interactions
can generate large amplitude waves, and that an initial state 
that is not an exact solution eventually converges to an exact 
multi-soliton solution. 
We note that resonant 2-soliton solutions with a hole have 
also been found in other discrete and continuous (2+1)-dimensional 
soliton equations \cite{infeld,yajima,JPA39p4063,Kenichi,tsuji1},
both in exact solutions and in numerical simulations,
indicating that they are a fundamental and robust
structure of (2+1)-dimensional soliton equations.

\section*{Acknowledgements}

We thank H. Segur for 
many interesting discussions.
KM also acknowledges partial support from the 21st Century 
COE program ``Development of Dynamic Mathematics with High Functionality'' 
at the Faculty of Mathematics, Kyushu University and Grant-in-Aid for 
Scientific Research from the Japan Society for the Promotion of Science. 

%

\eject


\bigskip
\begingroup
\small\sc\noindent
$^1$
State University of New York, Buffalo, NY, USA
\\[1ex]
$^2$ 
University of Texas-Pan American, Edinburg, TX, USA
\\[1ex]
$^3$
Research Institute for Applied Mechanics,
Kyushu University, Fukuoka, Japan
\endgroup

\bigskip
\noindent

Revised Version: Several discussions about amplitude were improved. 
Several figures were improved. 
Editorial production errors were corrected. 

\end{document}